\documentclass[twocolumn]{aastex62}
\usepackage{import}

\usepackage{mathtools}
\usepackage{chngcntr}
\usepackage{footnote}
\usepackage[caption=false]{subfig}
\usepackage{float}
\usepackage{tabulary,booktabs}

\usepackage{multirow}
\usepackage{graphicx}
\usepackage{amsmath}
\usepackage{yhmath}

\shorttitle{Environmental dependence of the MZR at $z=1.4-2.6$ }
\shortauthors{Chartab et al.}

\begin{document}

\title{\textbf{The MOSDEF Survey: Environmental dependence of the gas-phase metallicity of galaxies at $1.4 \leq z \leq 2.6$\footnote{Based on data obtained at the W.M. Keck Observatory, which is operated as a scientific partnership among the California Institute of Technology, the University of California, and NASA, and was made possible by the generous financial support of the W.M. Keck Foundation.}}}

\correspondingauthor{Nima Chartab}
\email{nima.chartab@email.ucr.edu}

\author[0000-0003-3691-937X]{Nima Chartab}

\affil{Department of Physics and Astronomy, University of California, Riverside, 900 University Ave, Riverside, CA 92521, USA}

\author{Bahram Mobasher}
\affiliation{Department of Physics and Astronomy, University of California, Riverside, 900 University Ave, Riverside, CA 92521, USA}

\author{Alice E. Shapley} \affiliation{Department of Physics \& Astronomy, University of California, Los Angeles, 430 Portola Plaza, Los Angeles, CA 90095, USA}

\author{Irene Shivaei} 
\altaffiliation{Hubble Fellow}
\affiliation{Department of Astronomy/Steward Observatory, 933 North Cherry Ave, Rm N204, Tucson, AZ, 85721-0065, USA}

\author{Ryan L. Sanders}
\altaffiliation{Hubble Fellow}
\affiliation{Department of Physics and Astronomy, University of California, Davis, One Shields Ave, Davis, CA 95616, USA}

\author{Alison L. Coil} \affiliation{Center for Astrophysics and Space Sciences, University of California, San Diego, 9500 Gilman Dr., La Jolla, CA 92093-0424, USA}

\author{Mariska Kriek} \affiliation{Astronomy Department, University of California, Berkeley, CA 94720, USA}

\author{Naveen A. Reddy} \affiliation{Department of Physics and Astronomy, University of California, Riverside, 900 University Ave, Riverside, CA 92521, USA}

\author{Brian Siana}
\affiliation{Department of Physics and Astronomy, University of California, Riverside, 900 University Ave, Riverside, CA 92521, USA}

\author{William R. Freeman}
\affiliation{Department of Physics and Astronomy, University of California, Riverside, 900 University Ave, Riverside, CA 92521, USA}

\author{Mojegan Azadi} \affiliation{Harvard-Smithsonian Center for Astrophysics, 60 Garden Street, Cambridge, MA 02138, USA}

\author{Guillermo Barro} \affiliation{Department of Physics, University of the Pacific, 3601 Pacific Ave, Stockton, CA 95211, USA}

\author{Tara Fetherolf}
\affiliation{Department of Physics and Astronomy, University of California, Riverside, 900 University Ave, Riverside, CA 92521, USA}

\author{Gene Leung}
\affiliation{Center for Astrophysics and Space Sciences, University of California, San Diego, 9500 Gilman Dr., La Jolla, CA 92093-0424, USA}

\author{Sedona H. Price} \affiliation{Max-Planck-Institut f{\"u}r extraterrestrische Physik, Postfach 1312, Garching, 85741, Germany}

\author{Tom Zick}
\affiliation{Astronomy Department, University of California, Berkeley, CA 94720, USA}

\begin{abstract}
\label{abstract}
Using the near-IR spectroscopy of the MOSFIRE Deep Evolution Field (MOSDEF) survey, we investigate the role of local environment in the gas-phase metallicity of galaxies. The local environment measurements are derived from accurate and uniformly calculated photometric redshifts with well-calibrated probability distributions. Based on rest-frame optical emission lines, $\rm[NII]\lambda6584$ and H$\alpha$, we measure gas-phase oxygen abundance of 167 galaxies at $1.37\leq z\leq1.7$ and 303 galaxies at $2.09\leq z\leq2.61$, located in diverse environments. We find that at $z\sim1.5$, the average metallicity of galaxies in overdensities with $ \rm M_*\sim10^{9.8}M_\odot, 10^{10.2}M_\odot\ and\ \rm10^{10.8}M_\odot$ is higher relative to their field counterparts by $0.094\pm0.051$, $0.068\pm0.028$ and $0.052\pm0.043$ dex, respectively. However, this metallicity enhancement does not exist at higher redshift, $z\sim2.3$, where, compared to the field galaxies, we find $0.056\pm0.043$, $0.056\pm0.028$ and $0.096\pm 0.034$ dex lower metallicity for galaxies in overdense environments with $\rm M_*\sim10^{9.8}M_\odot, 10^{10.2}M_\odot\ and\ 10^{10.7}M_\odot$, respectively. Our results suggest that, at $1.37\leq z\leq2.61$, the variation of mass-metallicity relation with local environment is small ($<0.1$dex), and reverses at $z\sim2$. Our results support the hypothesis that, at the early stages of cluster formation, owing to efficient gas cooling, galaxies residing in overdensities host a higher fraction of pristine gas with prominent primordial gas accretion, which lowers their gas-phase metallicity compared to their coeval field galaxies. However, as the Universe evolves to lower redshifts ($z\lesssim2$), the shock-heated gas in overdensities cannot cool down efficiently, and galaxies become metal-rich rapidly due to the suppression of pristine gas inflow and re-accretion of metal-enriched outflows in overdensities.     
\end{abstract}

\bigskip

\section{Introduction}
\label{Introduction}

Over the last few years, we have made significant progress towards developing a comprehensive and self-consistent model for the formation of galaxies. At each step, however, the model is compounded by non-linear effects regarding the feedback processes involved and the parameters deriving them. Gas accretion from the intergalactic medium (IGM) supplies cold gas for a galaxy to build up its stellar population through the gravitational collapse of molecular clouds. As the stars form, heavy elements are produced in their hot cores, resulting in the chemically enriched material which will be expelled to the interstellar medium (ISM) through feedback processes such as stellar winds \citep[][]{Garnett02,Brooks07} and/or supernovae explosions \citep[e.g.,][]{Steidel10,Martin12,Chisholm18}. Furthermore, feedback processes can remove part of the enriched material from galaxies into the IGM \citep[e.g.,][]{Heckman90,Tremonti04,Chisholm18}. Thus, gas-phase metallicity is expected to be connected with most of the galaxy evolutionary processes (e.g., gas inflow/outflow and star formation) and can be considered as one of the fundamental characteristics of galaxies that encodes information regarding galaxy evolution over cosmic time.

A tight correlation has been observed between the gas-phase metallicity and stellar mass of galaxies (Mass-Metallicity Relation, hereafter MZR) out to $z\sim 3.5$, such that galaxies with lower stellar masses have lower metallicities \citep[e.g.,][]{Tremonti04,Erb2006,Mannucci09,Finkelstein11,Steidel14,Sanders15,Sanders20}. Moreover, it has been found that the MZR evolves with redshift so that galaxies at a given stellar mass have lower metallicity at high redshifts \citep[e.g.,][]{Steidel14,Maiolino19,Sanders20}.

Although a tight correlation has been found between stellar mass and gas-phase metallicity spanning a wide range of stellar masses, other physical properties of galaxies could also play a role in contributing to the observed scatter in this relation. For example, the star formation rate (SFR) in galaxies controls their metallicity. At a given stellar mass, galaxies with lower SFR have higher metallicities \citep[e.g.,][]{Mannucci10,Sanders18}. Gas accretion from the IGM adds chemically poor gas into the ISM of galaxies, which lowers the galaxy's gas-phase metallicity. On the other hand, cold gas infall that provides additional fuel for star formation increases the SFR. Therefore, cold gas accretion changes the metallicity content of the ISM in a complicated way. \cite{Lilly13} introduced a gas regulator model that expresses the gas-phase metallicity of a galaxy in terms of the properties of the accreted gas (metallicity and rate of infalling gas) and SFR. Their model explains the observationally-confirmed dependence of the MZR on SFR. 

However, the evolution of the ISM and its metal content does, to a large extent, depend on the properties of the infalling gas which, in turn, is affected by the environment where the galaxy resides \citep{Peng14,Gupta18}. \cite{chartab19} studied the environmental dependence of star formation activity of galaxies in the five widely separated fields of the Cosmic Assembly Near-IR Deep Extragalactic Legacy Survey \citep[CANDELS;][]{2011ApJS..197...35G,2011ApJS..197...36K} out to $z\sim 3.5$. They found that environmental quenching efficiency evolves with stellar mass such that massive galaxies in overdense regions become quenched more efficiently than their low mass counterparts. This suggests that, besides the stellar mass quenching, which is the dominant quenching mechanism at high redshift, the environmental quenching is also effective as early as $z\sim 3$ for massive galaxies (i.e.,
the quiescent fraction of galaxies with $\rm M_*\sim 10^{11} M_\odot$ are $20\%$ higher \citep{chartab19} in overdensities than that of field counterparts at $z\sim 2.8$). They also suggest that the growth of environmental quenching efficiency with stellar mass can be explained by the termination of cold gas accretion in an overdense environment. In the absence of cold gas accretion, massive galaxies exhaust their remaining gas reservoirs in shorter timescales compared to low mass galaxies \citep[see also,][]{Balogh2016,Kawinwanichakij2017,Fossati17,old2020}. Using the Evolution and Assembly of GaLaxies and their Environments (EAGLE) simulation, \cite{vandeVooet2017} found a suppression of the cold gas accretion rate in dense environments, mostly for satellite galaxies. \cite{Zavala19} reported ALMA observation of 68 spectroscopically-confirmed galaxies within two proto-clusters at $z\sim 2.2$ and found that proto-clusters contain a higher fraction of massive and gas-poor galaxies compared to those residing in the field environment. They concluded that the environmental quenching exists during the early phases of cluster formation \citep[see also,][]{Darvish16,Nantais2017,Ji18,Pintos-Castro19,Contini2020,Ando2020}. However, \cite{Lemaux2020} recently observed an enhanced star formation for star-forming galaxies in overdensities at $z>2$ \citep[see also,][]{Elbaz07,Tran2010}. All these observations reveal the importance of the local environment in gas accretion rate and SFR that govern the gas-phase metallicity of galaxies.

Moreover, the metallicity of the infalling gas varies in different environments. In the local Universe, it is observed that the metallicity of IGM gas in cosmic voids is $<  0.02\ Z_{\odot}$ \citep{Stocke07}, while this number is $\sim0.3\ Z_{\odot}$ for a cluster-like environment \citep[e.g.,][]{Mushotzky97}. This is consistent with results from the IllustrisTNG simulations \citep{Gupta18} that predict a metal-enhanced infalling gas in a dense environment out to $z\sim 1.5$. Both pieces of evidence of lower gas accretion rate and higher metallicity of IGM gas in denser environments hint that a part of the scatter on the MZR could be due to the difference in the environment of galaxies. 

The metallicity of galaxies in diverse environments is extensively studied in the local Universe. Small but significant environmental dependence of the MZR is found, especially for satellite galaxies, such that at a given stellar mass, galaxies in overdensities have higher gas-phase metallicities \citep[e.g.,][]{Cooper08,Ellison2009,Peng14,Wu17,Schaefer19}. \cite{Cooper08} found that $\gtrsim 15\%$ of the measured scatter in the MZR is caused by environmental effects. Furthermore, \cite{Peng14} found that galaxies with high metallicities favor denser environments at $z\sim 0$. They conclude that higher metallicity of infalling gas in dense environments is responsible for the environmental dependence of the MZR.

Beyond the local Universe (at $z\gtrsim 1$), the situation is unclear. Some studies found evidence of enhanced gas-phase metallicity in low-mass cluster galaxies at $z\gtrsim 1.5$ \citep[e.g.,][]{Kulas13,Shimakawa15,Maier19}. \cite{Kulas13} used Keck/MOSFIRE observations to compare the gas-phase metallicity of 23 protocluster members ($z\sim 2.3$) with 20 field galaxies. They found that the mean metallicity of low mass galaxies ($\rm M_*\sim 10^{10} M_{\odot}$) in the protocluster is $\sim 0.15$ dex higher than that in the field. On the other hand, \cite{Valentino15} reported $0.25$ dex lower gas-phase metallicity for the members of a $z\sim 2$ protocluster ($\rm M_*\sim 10^{10.5} M_{\odot}$) compared to field galaxies at the same redshift. Conversely, some other studies have not observed significant environmental dependence of the MZR at high redshift \citep[e.g.,][]{Tran15,Kacprzak15,Namiki19}.

With the wealth of near-IR spectroscopy for the star-forming galaxies ($1.37\leq z\leq 2.61$) in the MOSFIRE Deep Evolution Field (MOSDEF) survey \citep{Kriek15}, combined with the local environment measurements \citep{chartab19} derived from accurate and uniformly calculated photometric redshifts with well-calibrated probability distributions, in this paper we investigate the effect of the local environment on the gas-phase metallicity of galaxies at $z\sim 1.5$ and $z\sim 2.3$. In Section \ref{sec:Data}, we present the details of the MOSDEF sample used in this work. We then briefly describe the local number density measurements, as a proxy for the environment, based on the photometric observations of the CANDELS fields. We describe sample selection procedure in Section \ref{sec:sample}. In Section \ref{environment_MZR}, we investigate the role of the environment in the MZR, followed by the physical interpretation of our observations. We discuss our results in Section \ref{sec:Discussion} and summarize them in Section \ref{sec:Summary}.

Throughout this work, we assume a flat $\Lambda$CDM cosmology with $H_0=100h \rm \ kms^{-1} Mpc^{-1}$, $\Omega_{m_{0}}=0.3$ and $\Omega_{\Lambda_{0}}=0.7$. All magnitudes are expressed in the AB system, and the physical parameters are measured assuming a \cite{cha03} IMF.


\section{Data}
\label{sec:Data}

\subsection{The MOSDEF survey} 
\label{sec:MOSDEF}
MOSDEF is an extensive near-IR spectroscopic survey conducted over 48.5 nights using the Keck/MOSFIRE spectrograph \citep{McLean12}. As a part of the survey, $\sim 1500$ galaxies in five CANDELS fields (EGS, COSMOS, GOODS-N, UDS, and GOODS-S) were observed in three redshift ranges where strong rest-optical emission lines fall within windows of atmospheric transmission ($400$ galaxies at $z\sim 1.5$, $750$ at $z\sim 2.3$, and $400$ at $z\sim 3.4$). These galaxies are selected from the 3D-HST photometric and spectroscopic catalogs \citep{Skelton14,Momcheva16} to a limiting HST/F160W magnitude of 24.0, 24.5, and 25.0 at $z\sim 1.5$, 2.3, and 3.4, respectively. Based on these magnitude limits, the MOSDEF sample is roughly mass-complete down to $\rm M_*\sim 10^{9.5} M_{\odot}$ \citep{Shivaei15a}. For a full description of the survey strategy, observation and data reduction, we refer readers to \cite{Kriek15}.

 The spectral energy distributions (SED) of MOSDEF galaxies are fitted using the multiwavelength photometric data from the 3D-HST survey \citep{Skelton14,Momcheva16}. Briefly, the photometric fluxes are corrected for contamination caused by strong nebular emission lines measured from the MOSDEF rest-frame optical spectra. The flexible stellar population synthesis model of \cite{con09} is adopted to build a library of synthetic spectral energy distributions. Star formation histories are modeled with a delayed exponentially declining function (${\rm SFR} \propto t e^{-t/\tau}$), where $t$ is the age of a galaxy and $\tau$ is the star formation timescale. Dust attenuation is applied using the \cite{Calzetti} law and solar stellar metallicity is assumed for all galaxies. Then, the SED fitting is performed using the FAST code \citep{kri09}, which uses a $\chi^2$ minimization method to find the best-fit stellar population model and corresponding properties such as stellar mass and SFR. Corresponding confidence intervals are computed by perturbing the photometry using the photometric errors. Redshifts are fixed to their spectroscopic values for the SED fitting. The spectroscopic redshifts and emission line fluxes are measured from the extracted 1D spectra, presented in \cite{Kriek15}.  

\subsection{Measuring local environment of galaxies} 
\label{sec:environment}
To quantify the environment of MOSDEF galaxies, we utilize the publicly available catalog of \cite{chartab19} which includes measurements of local density for galaxies brighter than $\rm HST/F160W \leq 26$ AB mag in all the five CANDELS fields: GOODS-S \citep{Guo13}, GOODS-N \citep{Barro2019}, COSMOS \citep{Nayyeri17}, EGS \citep{Stefanon17}, and UDS \citep{Galametz13}. 

As described in \cite{chartab19}, including both the spectroscopic and photometric redshifts for density measurements can bias the estimates in favor of galaxies with spectroscopic redshifts. Thus, despite the availability of spectroscopic redshifts for $\sim 12\%$ of galaxies in the CANDELS fields, the environment catalog relies on uniformly calculated photometric redshifts (normalized median absolute deviation $\rm \sigma_{NMAD}\sim 0.02$), with well-calibrated redshift probability distribution functions (PDF) \citep[][D. Kodra et al. in prep]{Kodra19}. The environment catalog has been constructed using the full photometric redshift PDFs adopting the technique of boundary-corrected weighted von Mises kernel density estimation \citep{chartab19}. Here we provide a brief explanation of density measurements.

\begin{figure}[!t]
\centering
\includegraphics[width=0.45\textwidth,clip=True, trim=5.5cm 1.5cm 5.5cm 1.5cm]{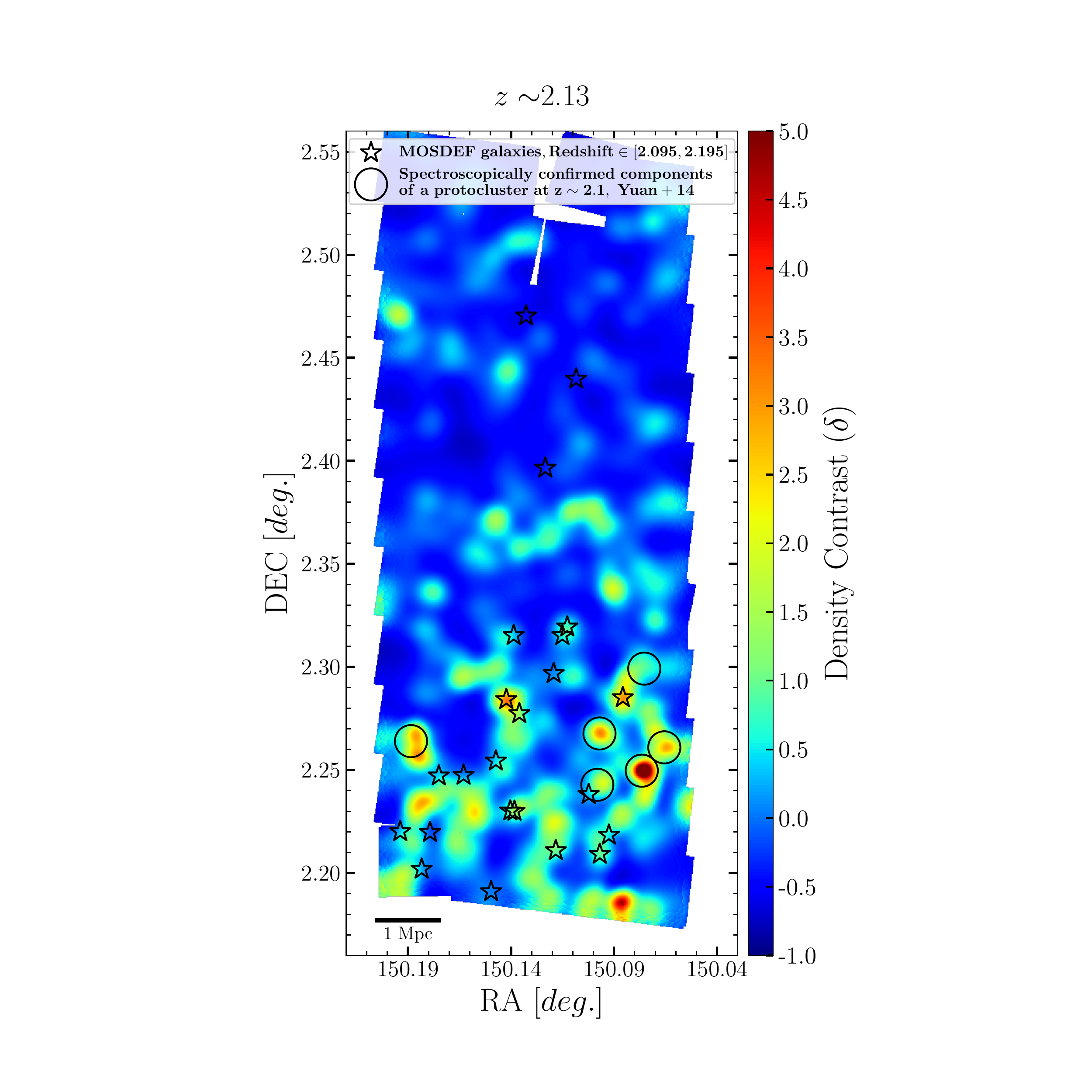}
\caption{An example of the density map for the CANDELS-COSMOS field at $z\sim 2.13$ \citep{chartab19}. Stars show MOSDEF sources located in this $z$-slice. \cite{Yuan14} spectroscopically confirmed six overdensities at this redshift which were initially identified by \cite{Spitler12}. They are denoted by open circles.}\label{Density map} 
\end{figure}

Taking advantage of the well-calibrated and tested photometric redshift PDFs of CANDELS galaxies, the position of galaxies are treated in a probabilistic way, such that all the information regarding the position of a galaxy in redshift space is embedded in its photometric redshift PDF. The dataset within the CANDELS fields is sliced to constant comoving width, $\Delta\chi=35 h^{-1} \text{Mpc}$ (e.g., $\Delta z=0.035$ at $z\sim 2$), which is greater than both redshift space distortion \citep[Fingers-of-God effect;][]{Jackson72} and the photometric redshift PDF resolution over the redshift range $0.4\leq z\leq 5$. Each galaxy is distributed over all redshift slices ($z$-slice) based on its photometric redshift PDF, such that the galaxy has a specific weight in each $z$-slice. Then, to calculate the local density in each $z$-slice, the weighted von Mises kernel density estimation is employed. The von Mises kernel is the spherical analog of the Gaussian kernel where variables are angles (e.g., right ascension and declination) instead of linear data \citep{Garcia13}. To quantify the environment, density contrast ($\delta$) is defined as

\begin{equation}
        \delta=\frac{\Sigma}{\Bar{\Sigma}}-1,
\end{equation} where $\Sigma$ is the number density of galaxies at the desired point and $\Bar{\Sigma}$ is the average density in the corresponding $z$-slice. Figure \ref{Density map} shows an example of the density map for one of the $z$-slices at $z\sim 2.13$ in the COSMOS field. One should note that the probabilistic nature of the method allows us to take into account the contribution of all galaxies (based on their photometric redshift PDFs) when we create density maps. In Figure \ref{Density map} we also include six spectroscopically confirmed overdensities at $z\sim 2.1$ \citep{Yuan14} which were initially discovered from the Magellan/FOURSTAR Galaxy
Evolution Survey (ZFOURGE) \citep{Spitler12}. We find that all six confirmed overdensities in \cite{Yuan14} are correctly predicted in the density map. 
To assign a density contrast for each galaxy, the weighted integration of the local densities over $z$-slices has been performed since a galaxy with a photometric redshift is not localized in redshift space but is distributed over all $z$-slices based on its PDF. The densities are also corrected for a systematic under-estimation near the edge of the survey footprint using the re-normalization technique.

In the following section, we cross-match the MOSDEF galaxies with the CANDELS photometric catalogs to find their associated local density measurements. Although we have spectroscopic redshifts for the MOSDEF galaxies, we do not pin them in density maps based on spectroscopic redshifts to measure their environments. The density measurements trace the relative position of galaxies in the survey and are not sensitive to systematic biases that may exist in the photometric redshifts, whereas measuring densities based on the photometric redshifts but defining positions of the MOSDEF galaxies in density maps by their spectroscopic redshifts leads to inconsistencies.


\section{Sample selection} 
\label{sec:sample}
We select MOSDEF galaxies that have $\rm H\alpha$ line luminosities with $\rm S/N\geq3$. We only include the star-forming galaxies based on the UVJ rest-frame color selection \citep{UVJ} derived from SED fitting. Objects that are flagged as an active galactic nucleus (AGN) in the MOSDEF catalog are excluded. The AGNs are identified based on X-ray emission or IRAC colors. We also require that $\rm log([NII]/H\alpha)<-0.3$ to exclude optical AGNs \citep{coi15,Azadi17,Azadi2018,Leung17,Leung19}. To have a mass-complete sample, we only consider galaxies with $\rm M_*\geq 10^{9.5} M_{\odot}$. These criteria result in a total of 560 galaxies at $0.78\leq z\leq 2.64$. We then cross-match these objects with the local environment catalog of \cite{chartab19} within a radius equal to the FWHM size of HST/F160w band point spread function, $\sim 0.2''$. For 23 galaxies, we could not find a source within the radius of $\sim 0.2''$ in the CANDELS photometry catalogs due to the difference in the source identification and photometry extraction between the CANDELS and the 3D-HST. Also, 15 galaxies were not included in the environment catalog since the catalog is constructed based on specific selection criteria: (1) SExtractor’s stellarity parameter $<0.95$, (2) requiring 95\% of photometric redshift PDF of the galaxy to fall within the redshift range of $0.4\leq z \leq 5$, and (3) brighter than 26 AB mag in HST/F160w band.
14 out of 15 missing galaxies in the environment catalog were incorrectly identified as low-z galaxies or had very broad photometric redshift PDFs which have not satisfied the second criterion, and the other missing galaxy was identified as a galaxy with HST/F160w $>26$ AB mag in the CANDELS photometric catalogs that has not satisfied the third criterion.

\begin{figure}
\centering
\includegraphics[width=0.5\textwidth]{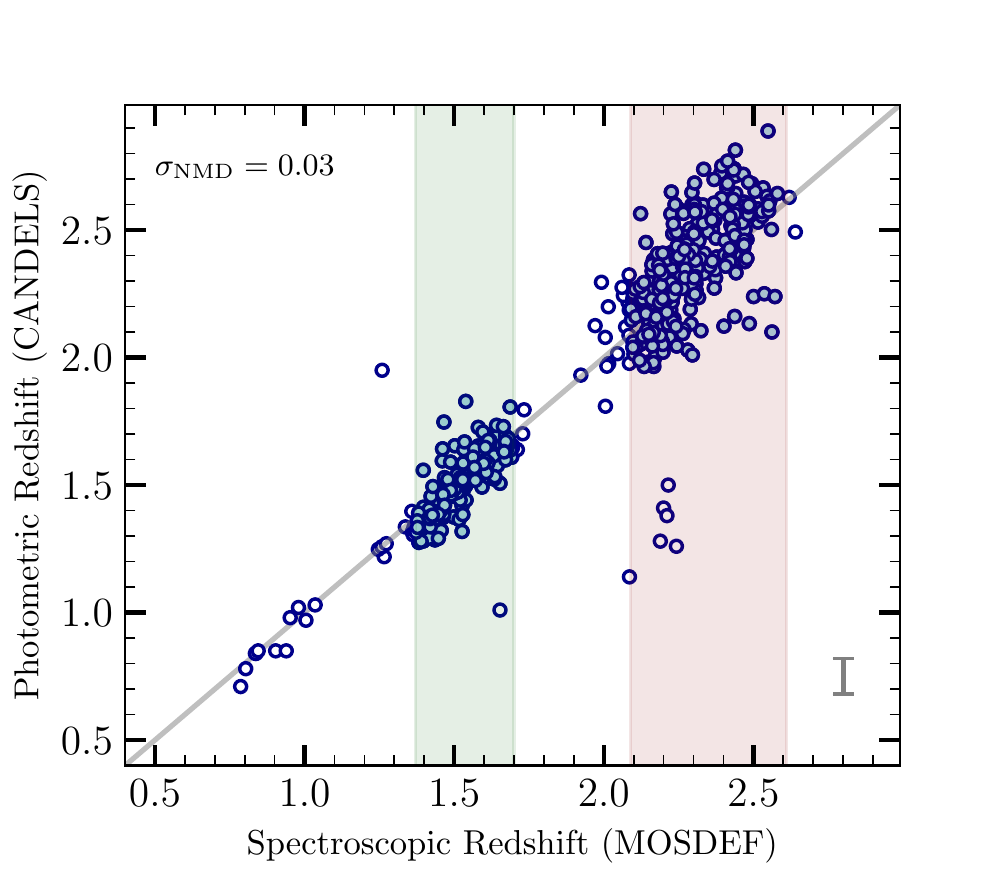}
\caption{Comparison between the MOSDEF spectroscopic and CANDELS photometric redshifts of the sample. The photometric redshift is defined as a probability-weighted expectation value of the redshift based on the photometric redshift PDF. The average uncertainty on photometric redshifts is displayed in the lower right corner. The normalized median absolute deviation of the redshift for the sample is 0.03. Solid blue circles show the final sample used in this work. The green shaded region corresponds to the lower redshift bin at $z\sim 1.5$, and the pink region shows the highest redshift bin at $z\sim 2.3$. Open circles are either outliers or galaxies outside the desired redshift ranges.}\label{z_comparison} 
\end{figure}

Figure \ref{z_comparison} shows the comparison between the spectroscopic and the CANDELS photometric redshifts of the sample. The photometric redshift is defined as a probability-weighted expectation value of the redshift based on the photometric redshift PDF \citep{Kodra19}. We find a value of 0.03 for the normalized median absolute deviation of photometric redshifts. This revalidates the accuracy of photometric redshifts, which results in reliable local density measurements. Furthermore, we removed outlier galaxies with the error $>0.5$ in their photometric redshifts and galaxies out of the desired redshift range (shown with open circles in Figure \ref{z_comparison}). The final sample is divided into two redshift bins, 167 galaxies at $1.37 \leq z \leq 1.70$, and 303 galaxies $2.09 \leq z \leq 2.61$.

\begin{figure*}
    \centering
	\includegraphics[width=1\textwidth]{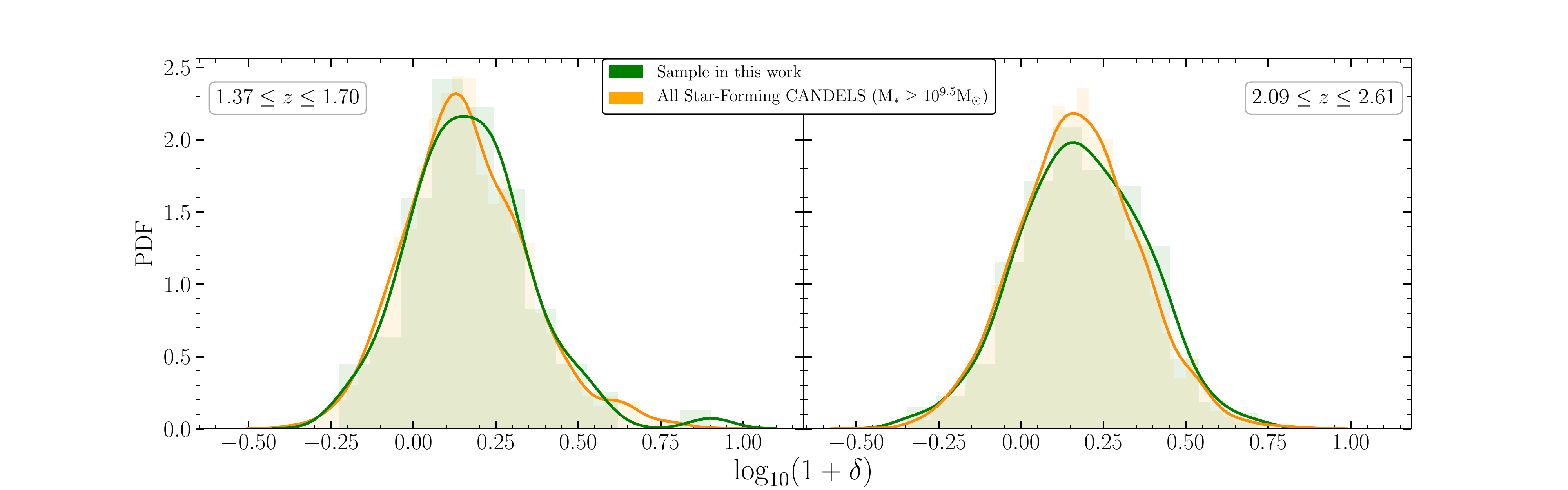}
	\caption{Histograms of overdensity measurements ($1+\delta$) for all the star-forming CANDELS galaxies with $\rm M_*\geq 10^{9.5} M_{\odot}$ and the sample used in this work at redshifts $z\sim 1.5$ (\textit{left}) and $z\sim 2.3$ (\textit{right}). Our sample covers a wide range of environments, making it unique to study the environmental dependence of spectroscopic properties of galaxies.}\label{overdensity_comparison} 
\end{figure*}

To investigate if our sample is representative of the full dynamic range of the local environment in the CANDELS data, we compare the local environmental density of CANDELS star-forming galaxies with $\rm M_*\geq 10^{9.5} M_{\odot}$ at the same redshift range with our sample (the MOSDEF galaxies). We use rest-frame UVJ colors computed in \cite{chartab19} to select a star-forming sub-sample of CANDELS galaxies. Figure \ref{overdensity_comparison} shows the histogram of density contrast for MOSDEF galaxies and for all of the star-forming CANDELS galaxies at the same redshift ranges. We find that MOSDEF galaxies cover a wide range of environments, making them unique for studying the environmental dependence of spectroscopic properties of galaxies. There is slight evidence that MOSDEF galaxies reside in relatively denser environments than the CANDELS star-forming sample which is expected for most spectroscopic surveys as they usually maximize the number of sources per mask. However, this effect is minimal in MOSDEF since it is an extensive program that covers $\sim 600\ \rm arcmin^2$ at $z\sim 2.3$ and $\sim 300\ \rm arcmin^2$ at $z\sim 1.5$ \citep{Kriek15}. Given the total area of the CANDELS fields ($\sim 960\ \rm arcmin^2$), MOSDEF covers $\sim 30\%$ and $\sim 60\%$ of the CANDELS area at $z\sim 1.5$ and $z\sim 2.3$, respectively. This wide areal coverage translates to diverse environments in our sample, as shown in Figure \ref{overdensity_comparison}.

We divide the sample into three bins of environment. Although we find that the local environment distribution of our sample is well representative of the full CANDELS galaxies, we do not set the environment binning thresholds based on our sample. Instead, we calculate them from all the star-forming CANDELS galaxies at desired redshift ranges. We find the tertiles which divide the CANDELS sample into three bins of environments, each containing a third of the sample. We consider galaxies in the lowest tertile ($\delta<\delta_{\frac{1}{3}}$) as field galaxies and those within the second tertile ($\delta_{\frac{1}{3}}\leq\delta<\delta_{\frac{2}{3}}$) as intermediate-densities and the highest tertile ($\delta\geq\delta_{\frac{2}{3}}$) as overdensities. Table \ref{environment-selected_sample} shows the density contrast thresholds to define the field, intermediate and overdense samples along with the sample size and the average density contrast, $\langle 1+\delta\rangle$ at each redshift bin. Our sample includes $\sim 55$ and $\sim 100$ galaxies in each bin of the environment at $z\sim 1.5$ and $z\sim 2.3$, respectively (see Table \ref{environment-selected_sample}).

\cite{Fossati15} linked observational local number density of galaxies to their parent halo masses using a stellar mass-limited sample of galaxies ($\rm M_*> 10^{9.5} M_{\odot}$) in semi-analytic models of galaxy formation. We estimate an average halo mass $\rm M_{halo} \rm \gtrsim 10^{13} M_{\odot}$, for overdensities (the highest tertile in present work) at $z\sim 2$ \citep[see also,][]{Fossati17}. Present-day descendants of these overdensities have $\rm M_{halo}\gtrsim 10^{14} M_{\odot}$ \citep{Behroozi13} associated with rich clusters. Massive core halos of these present-day rich cluster progenitors at $z\sim 2$ (protoclusters) have virial radii of $\lesssim 1$ comoving Mpc \citep{Chiang2017}, which can be observed within the CANDELS fields. Thus, our last environment bin, so-called overdensity, traces these massive cores of protoclusters at high redshift which will grow into $z=0$ clusters. However, all the dark matter and baryons that will assemble into a $z=0$ cluster may be very extended at $z\sim 2$, $\sim 50$ comoving Mpc \citep{Muldrew15}, which can not be captured in small CANDELS fields.    

We also estimate an average halo mass $\rm M_{halo}\lesssim 10^{12.5} M_{\odot}$ \citep{Fossati15} for the lowest tertile of the environment bins, so-called field galaxies. These halos will grow into halos with $\rm M_{halo}\lesssim 10^{13} M_{\odot}$ at $z=0$ \citep{Behroozi13}. Thus, field galaxies in the present work are the progenitor of galaxies residing in very poor $z=0$ groups/clusters (Local Group-like).  

In the following section, we model the MZR for our environment-selected samples to understand how metal enrichment processes of galaxies change with their respective environments.

\begin{table}[htp]
\centering
\caption{Properties of environment-selected sample}
\label{environment-selected_sample}
\begin{tabular}{cccc}
\hline
Environment & Sample size & $1+\delta$ & $\langle 1+\delta\rangle$ \\ \hline\hline
\multicolumn{4}{c}{$z\sim 1.5$} \\ \hline
Field & 53 & \textless{}1.19 & 0.97 \\ 
Intermediate-density & 54 & 1.19-1.69 & 1.43 \\ 
Overdense & 60 & \textgreater{}1.69 & 2.43 \\ \hline\hline
\multicolumn{4}{c}{$z\sim 2.3$} \\ \hline
Field & 97 & \textless{}1.24 & 0.97 \\ 
Intermediate-density & 96 & 1.24-1.77 & 1.49 \\ 
Overdense & 110 & \textgreater{}1.77 & 2.48 \\ \hline
\end{tabular}%
\end{table}


\section{Results}
\label{environment_MZR}

\subsection{The MZR in diverse environments}
\label{Average_MZR}
Here we use $\rm H\alpha$ and $\rm [NII]\lambda 6584$ lines to estimate oxygen abundances of galaxies as an indicator for their gas-phase metallicities. However, the $\rm [NII]\lambda 6584$ emission line is not detected with $\rm S/N\geq 3$ for 48 out of 167 and 118 out of 303 galaxies at $z\sim 1.5$ and $z\sim 2.3$, respectively. As we discuss later in Appendix \ref{appendix}, requiring $\rm [NII]\lambda 6584$-detection biases our sample toward higher gas-phase metallicities. Therefore, to include $\rm [NII]\lambda 6584$ non-detected galaxies in metallicity measurements, we create composite spectra by stacking the spectra of galaxies in bins of stellar mass and environment.

Following \cite{Shivaei18}, we shift individual spectra to the rest frame and then normalize them by the $\rm H\alpha$ luminosity. Composite spectra are computed by averaging the normalized spectra in bins of $0.5$ \AA\ considering the weights of $1/\sigma_i^2$ where $\sigma$ is the standard deviation of the $i^{th}$ spectra. The uncertainty in the weighted average is also obtained using $(\sum\frac{1}{\sigma_i^2})^{-\frac{1}{2}}$. 

The resultant composite spectra are normalized to the $\rm H\alpha$ luminosity. Therefore, the average flux ratio of $\rm \langle \frac{{[NII]\lambda 6584}}{{H{\alpha}}}\rangle $ is determined by fitting a triple Gaussian function for $\rm [NII]\lambda 6548,6584$ and $\rm H\alpha$ lines and extracting the area underneath the Gaussian function for the normalized $\rm {[NII]\lambda 6584}$ line. We perturb the composite spectra using their error distributions and estimate the average flux ratio for 500 trials. Also, galaxies in each bin of stellar mass and environment are bootstrap re-sampled in each trial to account for sample variance. The average and standard deviation of 500 trials are adopted as the flux ratio and its uncertainty.

We determine oxygen abundances ($\rm 12+\log(O/H)$) for the composite spectra using \cite{Pettini2004} calibration of the  $\rm N2=\frac{{[NII]\lambda 6584}}{H{\alpha}}$ line ratio,

\begin{equation}
    \rm 12+\log(O/H)=8.9+0.57\log( N2)
\end{equation}

The intrinsic uncertainty (1$\sigma$ dispersion) of the oxygen abundance of an individual galaxy in the calibration is 0.18. A composite spectrum consists of N galaxies; thus, the intrinsic error in the oxygen abundance of the composite spectrum is $0.18/\sqrt{N}$ \citep{Erb2006,Sanders15}. To include the N2 calibration error in measurements, one can calculate the total variance of metallicity by adding the intrinsic variance to the weighted average variance from the stacks. We do not include calibration error when reporting the uncertainty in metallicities. 
Moreover, the N2 estimator is calibrated locally and there is some debate within the literature regarding the validity of the calibration for high-redshift galaxies \citep{kew13,Steidel14,Shapley19,Sanders20}. However, it does not cause a problem for our work, where we study the relative metallicities to understand the effect of the local environment on the metal content of galaxies.

\begin{figure*}%
    \centering
    \subfloat{{\includegraphics[width=8.5cm,clip=True, trim=0.6cm 0.1cm 1.2cm 0.75cm]{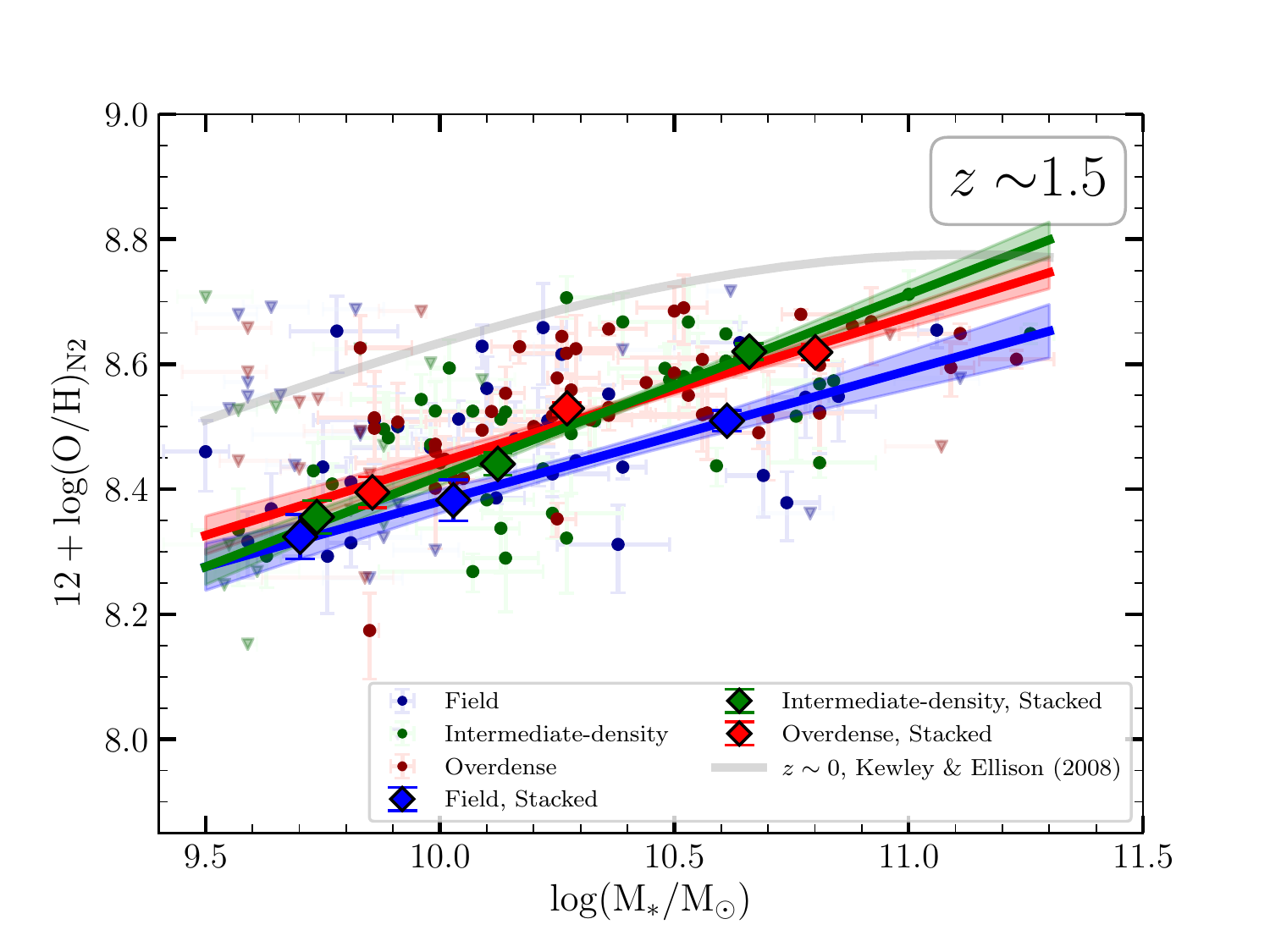} }}%
    \qquad
    \subfloat{{\includegraphics[width=8.5cm, trim=0.6cm 0.1cm 1.5cm 0.75cm]{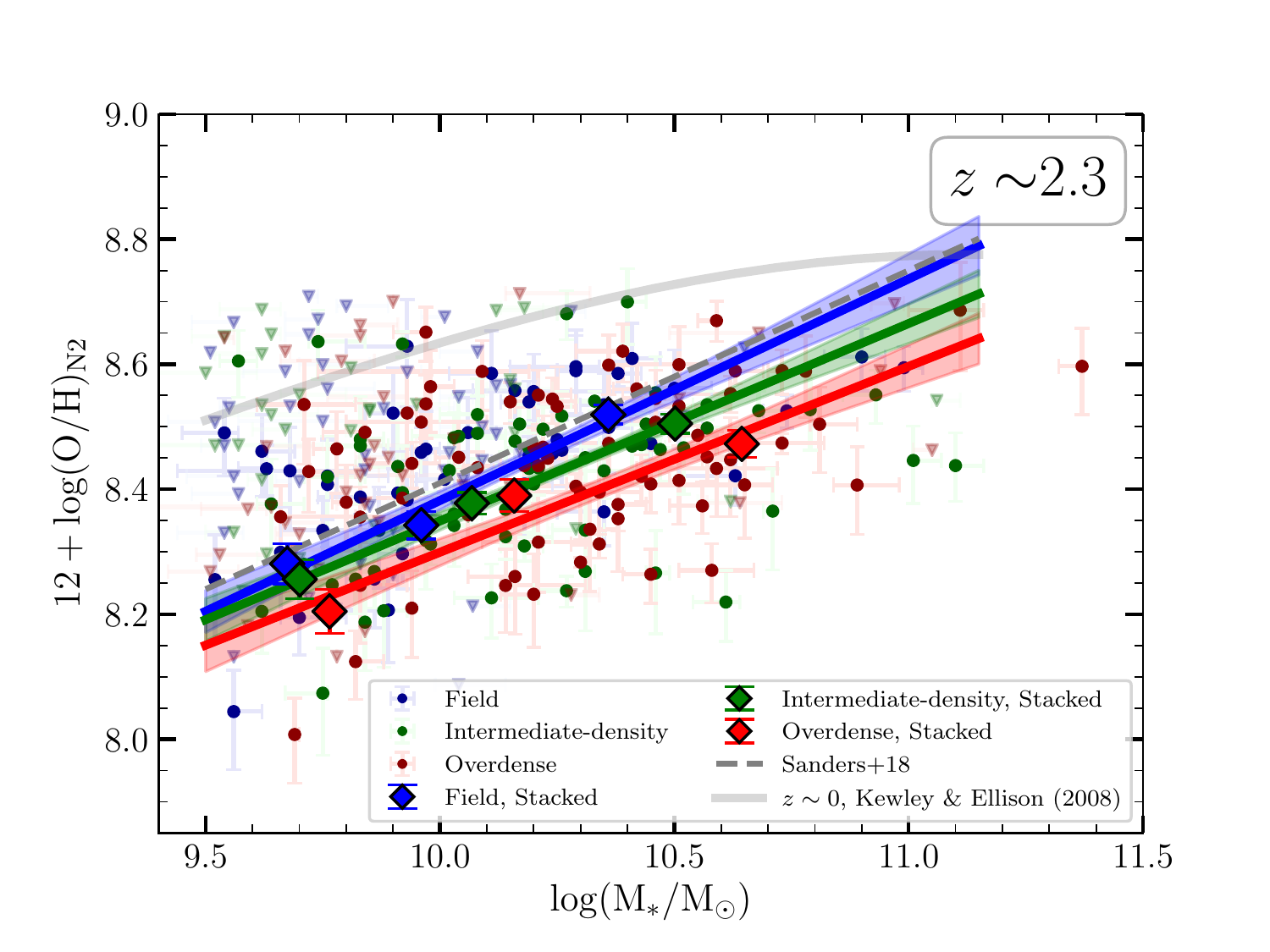} }}%
    \caption{MZR at $z\sim 1.5$ (\textit{left}) and at $z\sim 2.3$ (\textit{right}) for galaxies residing in 3 different environments: Overdensity (red), Intermediate-density (green), and underdensity (blue). For $\rm [NII]\lambda 6584$ non-detection, the upper limit of the metallicity is shown with inverted triangle symbols. The metallicity measurements for the composite spectra are shown with diamonds. The fitted average MZR lines (based on Table \ref{best_param}) for different environments are shown along with the $1\sigma$ error bars (the shaded regions around the best-fit models). The best-fit models suggest that the metallicity at a given stellar mass enhances in denser environments at $z\sim 1.5$ while the trend reverses at $z\sim 2.3$ such that galaxies in denser environments tend to have lower metallicities. Errors in the metallicities of composite spectra are not shown as they are smaller than the symbols (diamond). The SDSS local MZR \citep{Kewley08} is also shown by solid grey line and the average MZR for all the MOSDEF galaxies at $z\sim 2.3$ from \cite{Sanders18} is demonstrated by dashed grey line.}%
    \label{fig:MZR_fit}%
\end{figure*}

We further separate our sample in different environments into three bins of stellar mass such that each bin includes approximately an equal number of galaxies. We then calculate composite line luminosities for these nine bins of environments and stellar masses and calculate composite metallicities. To determine the average MZR for each environment (field, intermediate-density, and overdense), we fit a linear function to the composite metallicities and average stellar masses,
\begin{equation}
\label{MZR_model}
    \rm 12+\log(O/H)=\beta+\alpha \log(\frac{M_*}{M_\odot}),
\end{equation} where $\alpha$ and $\beta$ are the slope and intercept of the linear function, respectively. The best-fit parameters and errors are given in Table \ref{best_param}. To estimate errors, we perturb metallicities according to their uncertainties and repeat the fitting process 500 times. We also include best-fit parameters from \cite{Sanders18} for the MOSDEF galaxies at $z\sim 2.3$. Their sample is the same as ours except that they have one more constraint on the detection ($\rm S/N>3$) of the $\rm H\beta$ emission line.

\begin{table}[!h]
\centering

\caption{Best-fit linear parameters for MZR\footnote{Best-fit parameters for $\rm 12+\log(O/H)=\beta+\alpha\log(M_*/M_\odot)$}}
\label{best_param}
\begin{tabular}{ccc}
\hline
Environment & $\alpha$ & $\beta$\\ \hline\hline
\multicolumn{3}{c}{$z\sim 1.5$} \\ \hline
Field & $0.21\pm 0.04$ & $6.29\pm 0.43$ \\ 
Intermediate-density & $0.29\pm 0.03$ & $5.51\pm 0.31$ \\ 
overdense & $0.23\pm 0.03$ & $6.11\pm 0.31$ \\ \hline\hline
\multicolumn{3}{c}{$z\sim 2.3$} \\ \hline
Field & $0.35\pm 0.05$ & $4.84\pm 0.48$ \\ 
Intermediate-density & $0.31\pm 0.04$ & $5.19\pm 0.44$ \\ 
overdense & $0.30\pm 0.04$ & $5.33\pm 0.49$\\ \hline\hline
Sanders+18 \footnote{Best-fit parameters from \cite{Sanders18} for MOSDEF galaxies at $z\sim 2.3$ without any constraint on their environments.} & $0.34$ & $5.01$ \\ \hline
\end{tabular}
\end{table}

The resulting MZRs in different environments are shown in Figure \ref{fig:MZR_fit} along with the metallicity and stellar mass of individual galaxies. The metallicity measurements for the composite spectra are shown with diamonds. We find an enhancement in the gas-phase metallicity of galaxies in dense environments at $z\sim 1.5$ compared to the field galaxies, while the trend reverses at $z\sim 2.3$ such that galaxies in the dense environment tend to have lower metallicities compared to their counterparts in lower-density environments.

At $z\sim 1.5$, the average galaxies with $\rm 10^{9.5} M_\odot \lesssim M_* \lesssim 10^{11} M_\odot$ which reside in overdensities have enhanced metallicities by $\sim 0.07$ dex compared to their field counterparts. At the low stellar mass end of the MZR, this enhancement is less significant, mostly due to a higher fraction of $\rm [NII]\lambda 6584$ non-detection galaxies in the low mass end. At $z\sim 2.3$, the average galaxies with $\rm 10^{9.5} M_\odot \lesssim M_* \lesssim 10^{11} M_\odot$ in the dense environment are $\sim 0.11$ dex metal deficient relative to the field galaxies at the same stellar mass range.

A notable caveat for studies with a small sample size, which is the case for most of the high-redshift spectroscopic samples, is dissimilar stellar mass distributions among the galaxies in different environments. It is crucial to have the same stellar mass distributions in different environments to properly study the effect of the environment on the gas-phase metallicity of galaxies at a given stellar mass. As we bin the data to find the average MZR in diverse environments, different stellar mass distributions can result in different average metallicities. This difference can be misinterpreted as an environmental imprint on the MZR. Therefore, before proceeding to a discussion of our results, in the next section, we first perform an analysis on a mass-controlled sample of galaxies where we carefully match the stellar mass distribution of galaxies in different environments and find the composite spectra for the mass-controlled sample. This allows us to properly disentangle the effect of stellar mass from the local environment, providing an unbiased measurement of environmental trend.

\begin{figure*}
    \centering
	\includegraphics[width=1\textwidth,clip=True, trim=4cm 0cm 4cm 0cm]{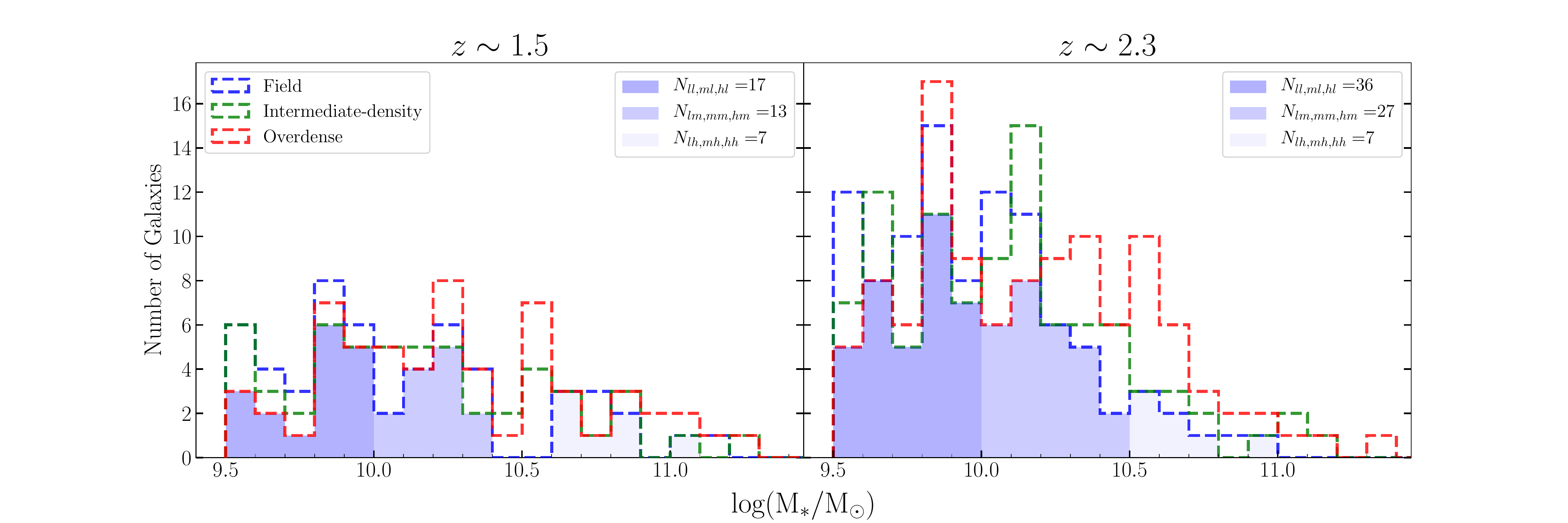}
	\caption{Histograms of stellar mass for the sample in different environments (dashed lines). Shaded regions show a sub-sampled data with the matched stellar mass distributions. At a given stellar mass, sub-sampled data have the same number of galaxies in every environment bin. The number of galaxies in each sub-sampled bin is shown in the format of $N_{xy}$, where $x$ shows the type of environment and $y$ shows the stellar mass range ("$l$": low, "$m$": intermediate and "$h$": high). For example, $N_{lh}=7$ at $z\sim 1.5$ means that each sub-sampled data consists of 7 massive field galaxies at that redshift.}\label{Mass_hist} 
\end{figure*}

\subsection{Mass-controlled sample}
\label{Mass-controlled sample}
The stellar mass distribution of our sample is shown in Figure \ref{Mass_hist}. As expected, galaxies in diverse environments have different stellar mass distributions. We find that overdensities tend to have a higher fraction of massive galaxies compared to underdense regions. For instance, the median stellar mass of our field sample at $z\sim 1.5$ is $\rm 10^{9.9} M_\odot$, while this value for the overdense sample is $\rm 10^{10.25} M_\odot$. In the highest redshift bin, $z\sim 2.3$, the median stellar masses are $\rm 10^{9.93} M_\odot$ and $\rm 10^{10.17} M_\odot$ for field and overdense galaxies, respectively. It is, therefore, important to match the stellar mass distribution of galaxies in different environments to properly disentangle the effect of the local environment on the gas-phase metallicity from the stellar mass. As shown in Figure \ref{Mass_hist}, we match the stellar mass distribution of galaxies in three environment bins by sub-sampling the galaxies such that each environment has the same number of galaxies at a given stellar mass. We adopt a resolution of $\rm \log(M_*/M_\odot)=0.1$ dex when matching the stellar mass distributions. In other words, we draw a fraction of galaxies in different environments such that they have the same stellar mass distributions with 0.1 dex resolution. As there are not unique mass-matched sub-samples in different environments, we repeat sub-sampling 500 times. For each trial, we measure the composite spectra in the bins of environment and stellar mass using the same procedure described in Section \ref{Average_MZR}. The stellar mass bins of $\rm 9.5\leq\log(M_*/M\odot)< 10$, $\rm 10\leq\log(M_*/M\odot)< 10.5$, and $\rm \log(M_*/M\odot)\geq 10.5$ are adopted. We perturb the resultant composite spectra of each trial according to their uncertainties. Ultimately, we construct the final mass-controlled composite spectra and their errors, using the average and standard deviation of the 500 measurements, respectively.    

We do not take into account stellar mass uncertainties when matching the stellar mass distribution of galaxies in different environments, since the median uncertainty of stellar masses in our sample is $\sim 0.05$ dex, which is smaller than the desired resolution in the mass-controlled sample, $0.1$ dex. 

\subsubsection{Metallicity of mass-controlled sample}

\begin{figure*}
\centering  
{\large (a) Composite spectra: \Large $z\sim 1.5$}
\subfloat{\includegraphics[width=1\linewidth,clip=True, trim=2.5cm 0cm 5cm 1.7cm]{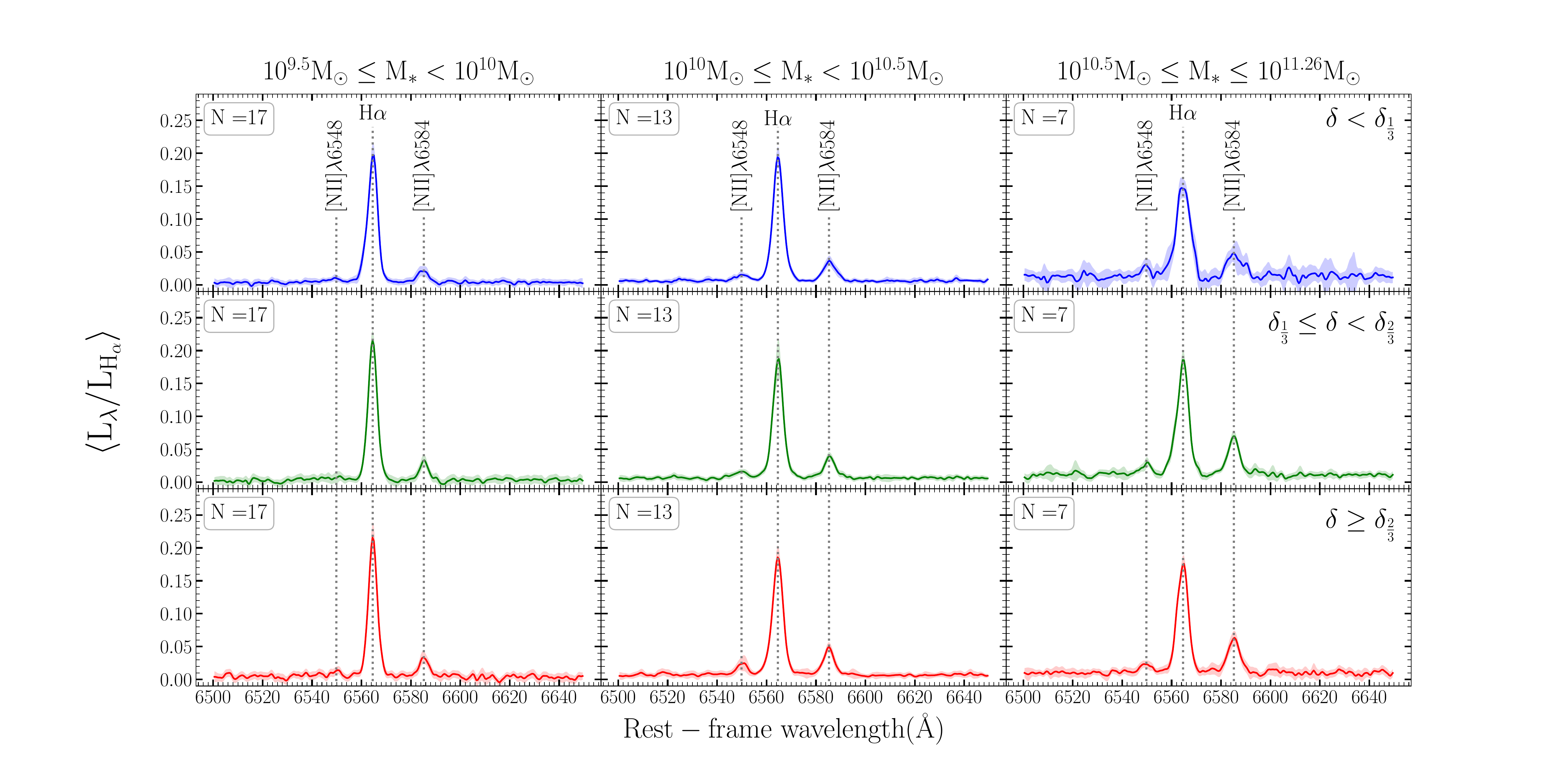}}
\qquad
\centering  

{\large (b) Composite spectra: \Large $z\sim 2.3$}
\subfloat{\includegraphics[width=1\linewidth,clip=True, trim=2.5cm 0cm 5cm 1.7cm]{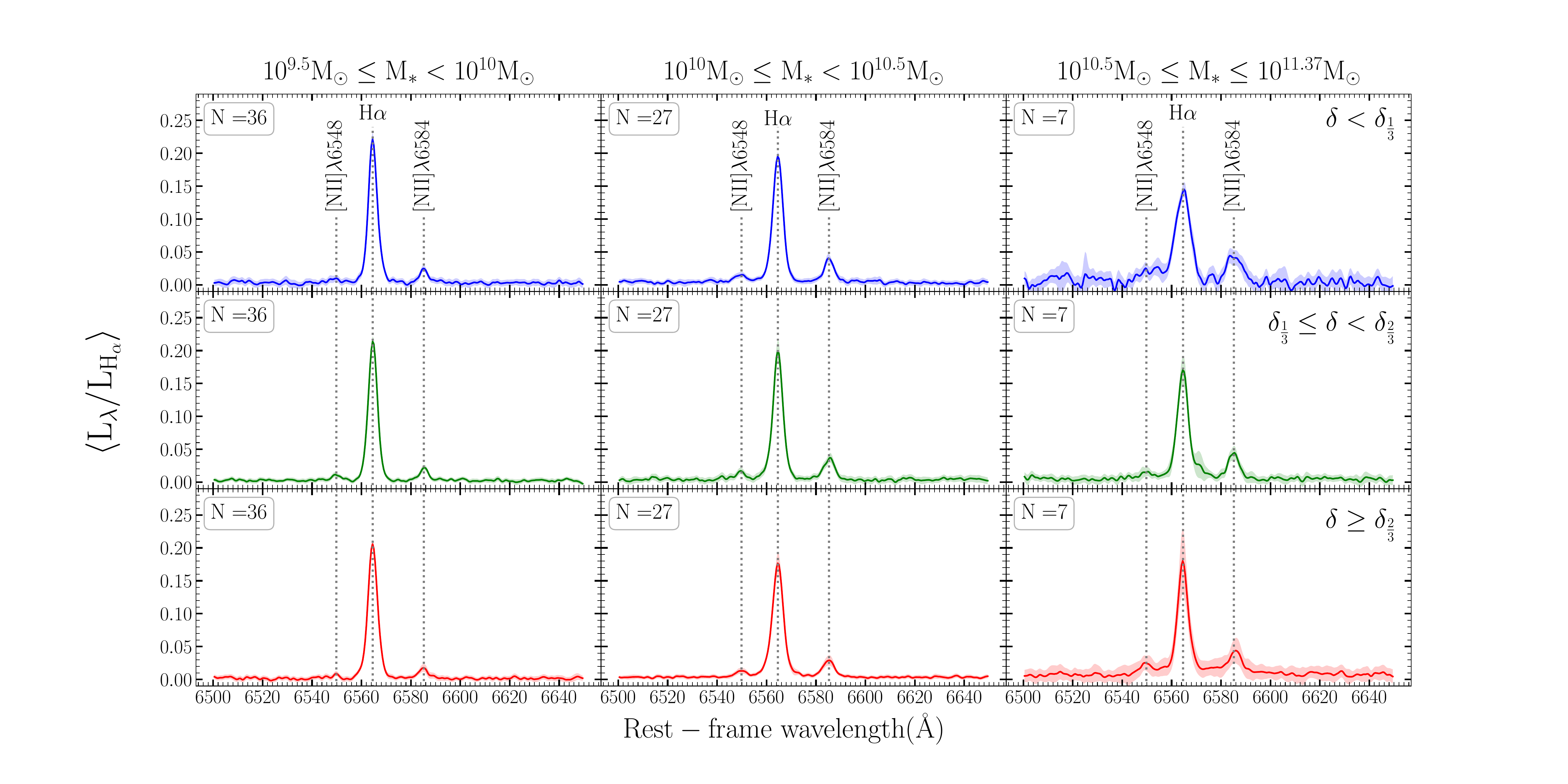}}
\qquad
\caption{Composite spectra for the mass-controlled samples in the bins of stellar mass and environment at $z\sim 1.5$ (\textit{top}) and $z\sim 2.3$ (\textit{bottom}). Errors are represented with the shaded regions, which are around the weighted average spectra. To build the composite spectra for the mass-controlled samples, we use the sub-sampling technique. Every time we sub-sample the data such that they have similar stellar mass distributions in three bins of environment, we build the composite spectra and perturb them according to their uncertainties. In the end, using the average and standard deviation of 500 trials, we construct mass-controlled composite spectra and their errors, respectively.}
\label{composite spectra}
\end{figure*}

The composite spectra for the nine bins of stellar masses and environments for the mass-controlled sample are shown in Figure \ref{composite spectra}. The $\rm \frac{{[NII]\lambda 6584}}{{H{\alpha}}}$ estimates and gas-phase metallicity (oxygen abundance) for mass-controlled composite spectra in the nine bins of stellar masses and environments are presented in Table \ref{table:composite spectra measurments}. We follow the same procedure as described in Section \ref{Average_MZR} to measure the metallicity of galaxies in the bins of environment and stellar mass from the composite spectra using the N2 indicator. 

We also measure the average stellar masses and SFRs of the mass-matched samples in stellar mass and environment bins. We use SFRs derived from SED fitting since the dust-corrected $\rm H\alpha$ luminosity is not available for $40\%$ of our sample. We avoid imposing a constraint on the detection of the H$\rm \beta$ emission line needed for dust correction, as it decreases our sample size significantly. The variation of metallicity at fixed $\rm M_*$ in different environments is small ($<0.1$ dex) and thus requires large sample sizes to be detected. Previous studies found that the SED-derived SFRs for MOSDEF galaxies are in general agreement with SFRs derived from dust-corrected $\rm H\alpha$ luminosities \citep{Shivaei16,Reddy15}. The average SED-derived SFRs listed in Table \ref{table:composite spectra measurments} suggest that even though there is evidence for a weak environmental dependence of the SFRs for our star-forming sample at a given stellar mass, it is not significant due to large uncertainties in the SFR measurements. Errors of SFRs listed in Table \ref{table:composite spectra measurments}  include uncertainty in the SFR of individual galaxies as well as sample variance.     A detailed study of the environmental imprints on specific SFRs of MOSDEF galaxies can properly constrain this relation. \cite{old2020} have used $\rm [OII]$-derived SFRs to investigate this relation at $1.0 < z < 1.5$ in Gemini Observations of
Galaxies in Rich Early Environments Survey \citep[GOGREEN;][]{Balogh17}. They find no significant difference between the specific SFR of the cluster and the field sample at $z>1.3$.

\begin{figure*}[t!]
\label{MZR_mass_matched}
    \centering
    \subfloat{{\includegraphics[width=8.5cm,clip=True, trim=0.6cm 0.2cm 1.5cm 0.75cm]{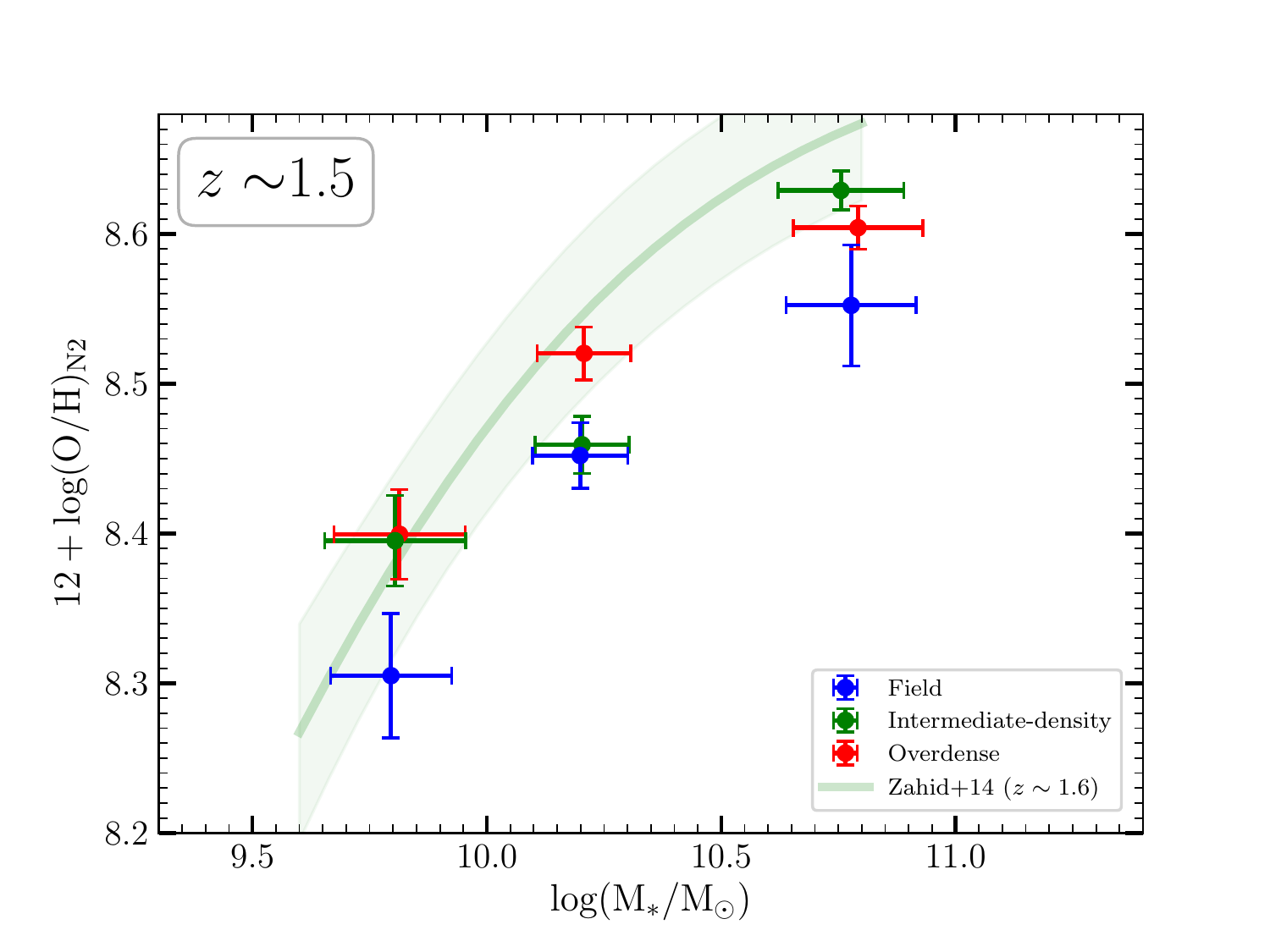} }}%
    \qquad
    \subfloat{{\includegraphics[width=8.5cm, trim=0.6cm 0.2cm 1.5cm 0.75cm]{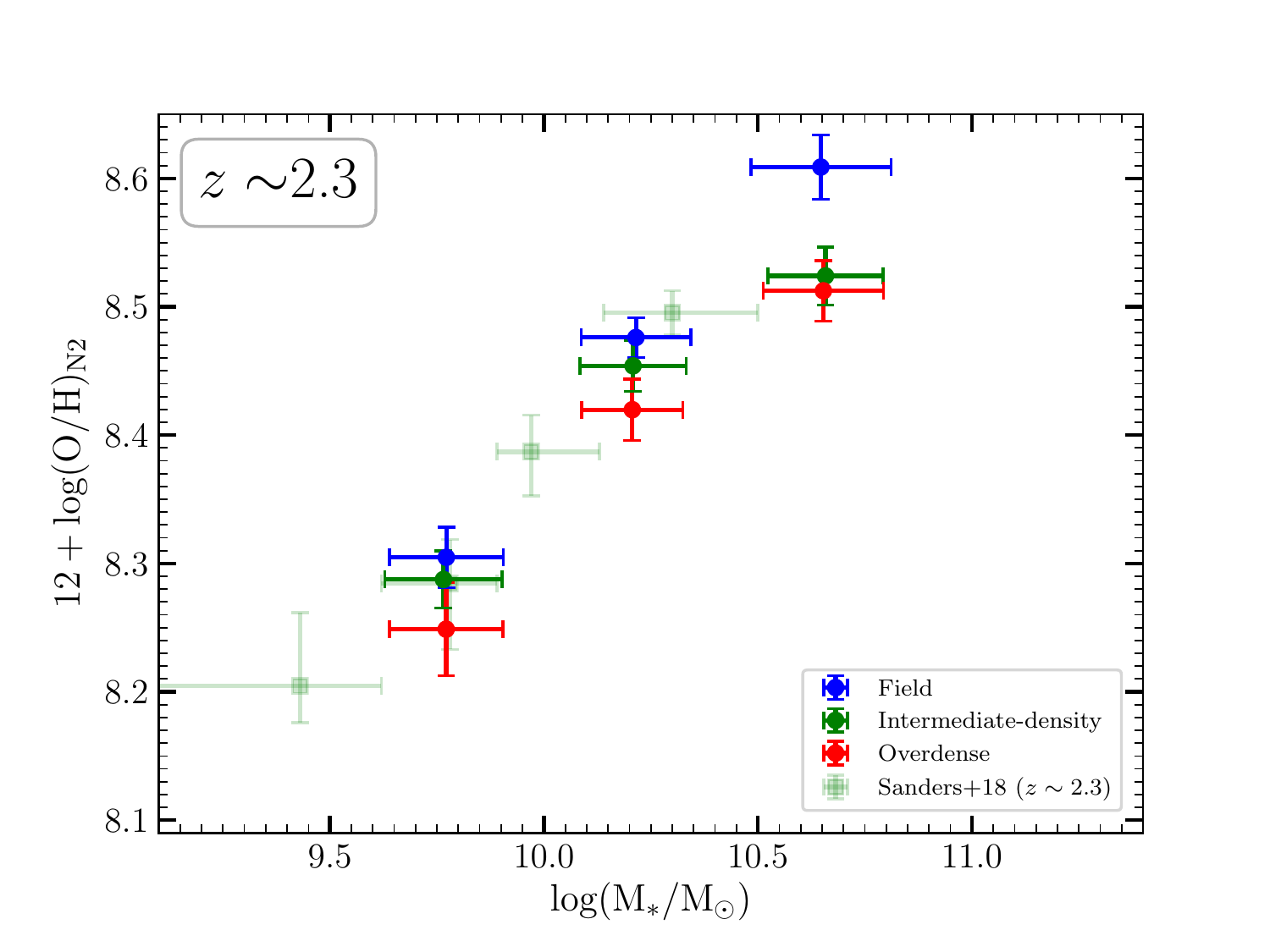} }}%
    \qquad
    \subfloat{{\includegraphics[width=8.5cm, trim=0.6cm 0.1cm 1.5cm 0.75cm]{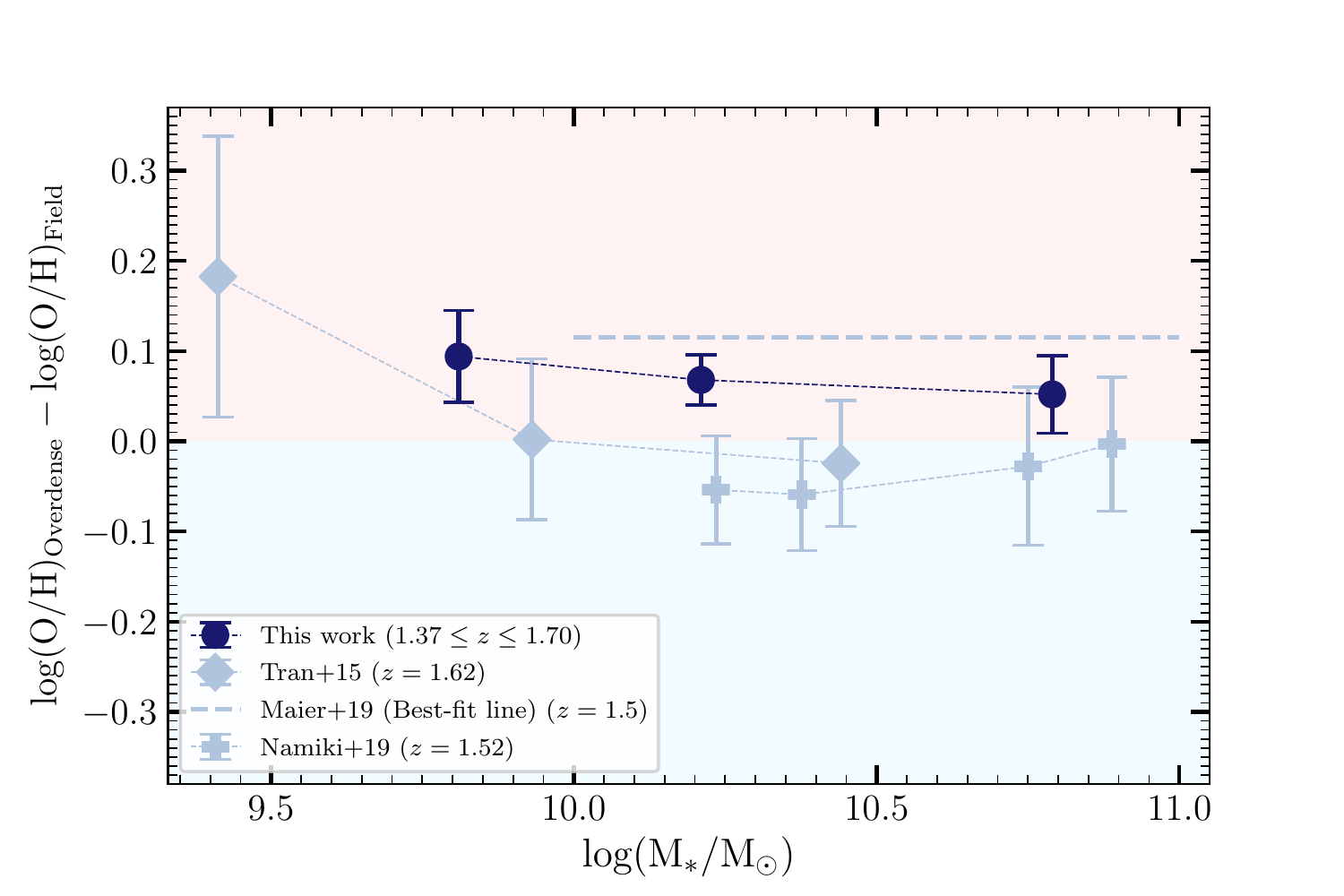} }}%
    \qquad
    \subfloat{{\includegraphics[width=8.5cm, trim=0.6cm 0.1cm 1.5cm 0.75cm]{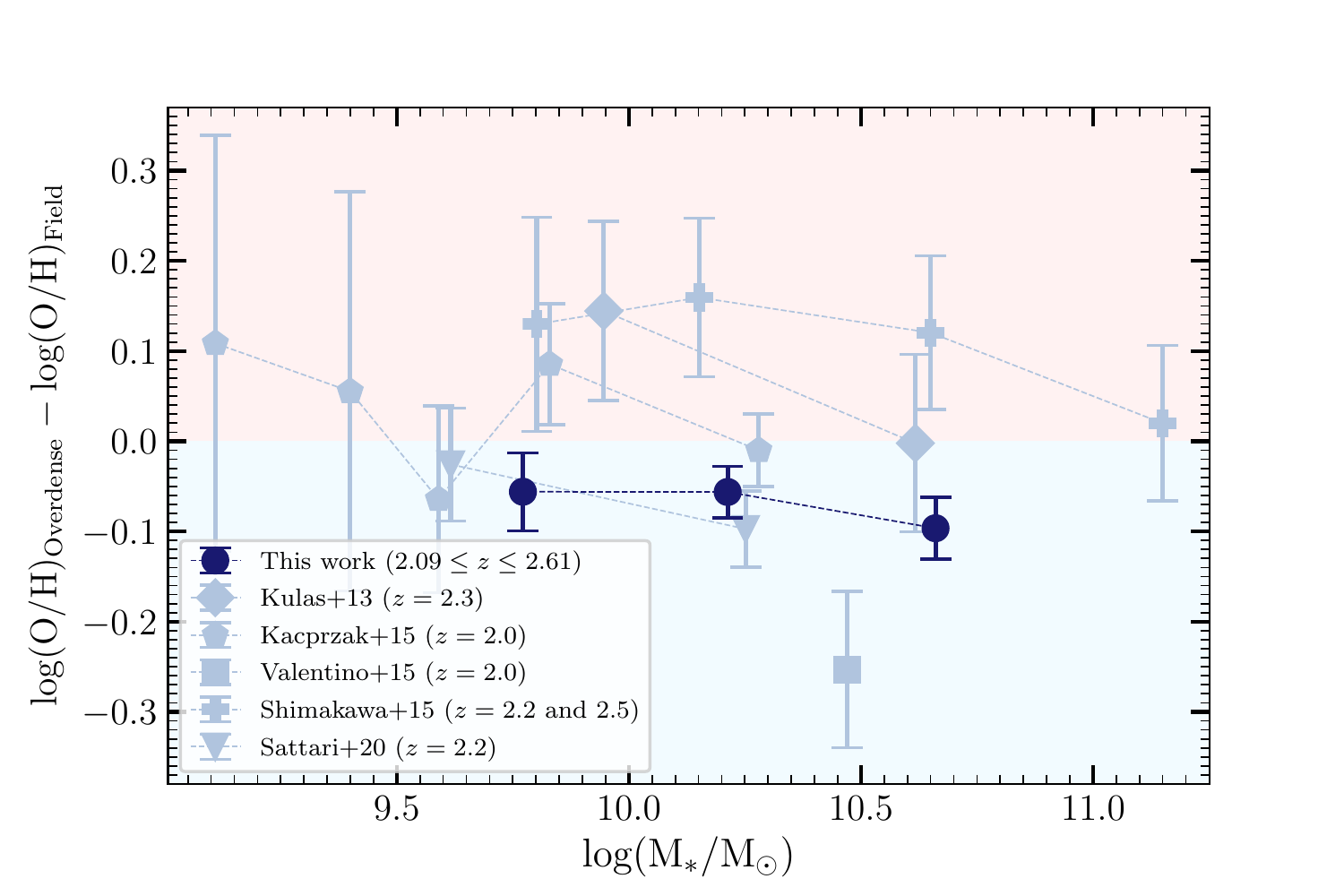} }}%
    \caption{Top panels: MZR for mass-matched sample at $z\sim 1.5$ ({\it left}) and $z\sim 2.3$ ({\it right}) in three different environments: Field (blue), Intermediate-density (green) and Overdense (red). Three stellar mass bins are fixed at $\rm 9.5\leq\log(M_*/M\odot)< 10$, $\rm 10\leq\log(M_*/M\odot)< 10.5$ and $\rm \log(M_*/M\odot)\geq 10.5$ and error bars in stellar masses show the $1\sigma$ scatter of the data around the average value. Bottom panels: The offset between the average metallicity of galaxies in overdensities (protocluster/cluster) and that of field galaxies as a function of stellar mass at $z\sim 1.5$ ({\it left}) and $z\sim 2.3$ ({\it right}). For comparison, previous studies in the literature at both redshifts are also included in the figures (see Section \ref{Comparison with previous works} for details).}%
    \label{MZR_mass_controlled}%
\end{figure*}

The MZR for the mass-controlled sample at $z\sim 1.5$ and $z\sim 2.3$ are presented in Figure \ref{MZR_mass_controlled}. We emphasize that the stellar mass distributions in the environment bins are similar. This allows us to remove the effect of stellar mass on metallicity to properly investigate the environmental effects. Also, although the mass-matched sample has fewer number of galaxies than the whole sample, we sub-sample the data 500 times to incorporate the contribution of the full sample in measurements. As shown in Figure \ref{MZR_mass_controlled}, we find that at a given stellar mass, the metallicity of galaxies changes with their respective environments at both redshift intervals considered here. 

At $z\sim 1.5$, the average metallicity of galaxies in overdensities is higher than that of field galaxies, with enhancements of $0.094\pm 0.051$ (1.8$\sigma$ significance), $0.068\pm 0.028$ (2.4$\sigma$ significance) and $0.052\pm 0.043$ (1.2$\sigma$ significance) dex for the mass-controlled sample with $ \rm M_*\sim 10^{9.8} M_\odot, 10^{10.2} M_\odot$ and $\rm 10^{10.8} M_\odot$, respectively. For galaxies residing in the intermediate-density, the metallicity enhancements are $0.090\pm 0.052$ (1.7$\sigma$ significance), $0.007\pm 0.029$ (insignificant) and $0.077\pm 0.042$ (1.8$\sigma$ significance) dex at the same stellar masses mentioned above.
 
In contrast, at $z\sim 2.3$, the average metallicity of galaxies in overdensities is lower than their field counterparts. Galaxies in the mass-controlled sample that reside in overdensities with $ \rm M_* \sim 10^{9.8} M_\odot, 10^{10.2} M_\odot$ and $\rm 10^{10.7} M_\odot$, are metal deficient by $0.056\pm 0.043$ (1.3$\sigma$ significance), $0.056\pm 0.028$ (2$\sigma$ significance) and $0.096\pm 0.034$ (2.8$\sigma$ significance) dex relative to the field sample, respectively. At this redshift, the metal deficiencies are $0.017\pm 0.032$ (insignificant), $0.022\pm 0.025$ (insignificant) and $0.085\pm 0.034$ (2.5$\sigma$ significance) dex for galaxies located in the intermediate-density with the same stellar masses as above.
\begin{table*}
\caption{Properties of the composite spectra for the mass-controlled sample}
\centering 

\label{table:composite spectra measurments}
\begin{tabular}{|c|c|cccccc|}
\hline
Redshift & Environment& $\rm N_T$ \footnote{Total number of galaxies in the bins of environment and stellar mass which is sub-sampled 500 times.} & $\rm N$ \footnote{Number of galaxies in each sub-sample (shaded region in Figure \ref{Mass_hist}).} &$\rm \langle \log{\frac{M_*}{M_\odot}}\rangle$ & $\rm \langle \log({SFR [{M_\odot\over year}]})\rangle$ & $\rm \langle N2 \rangle= \langle \frac{{[NII]\lambda 6584}}{{H{\alpha}}}\rangle $ & $\rm \langle 12+\log(O/H) \rangle$\\
\hline\hline

\multirow{9}{*}{1.37--1.70} & \multirow{3}{*}{Field} 
&27&17&9.79$\pm$0.03&0.89$\pm$0.24&0.090$\pm$0.015&8.305$\pm$0.042\\ &
&16&13&10.20$\pm$0.03&1.15$\pm$0.23&0.164$\pm$0.014&8.452$\pm$0.022\\ &
&9&7&10.78$\pm$0.05&1.15$\pm$0.35&0.246$\pm$0.040&8.552$\pm$0.040\\

\cline{2-8} 
                  & \multirow{3}{*}{Intermediate-density} 
                  &22&17&9.81$\pm$0.04&0.75$\pm$0.18&0.130$\pm$0.016&8.395$\pm$0.030\\ &
&21&13&10.20$\pm$0.03&1.08$\pm$0.32&0.169$\pm$0.013&8.459$\pm$0.019\\ &
&7&7&10.75$\pm$0.05&1.31$\pm$0.36&0.335$\pm$0.018&8.629$\pm$0.013\\

\cline{2-8} 
                  & \multirow{3}{*}{Overdense} 
                  &18&17&9.81$\pm$0.03&0.78$\pm$0.20&0.132$\pm$0.016&8.400$\pm$0.030\\ &
&17&13&10.21$\pm$0.03&1.26$\pm$0.29&0.216$\pm$0.015&8.520$\pm$0.018\\ &
&9&7&10.79$\pm$0.05&1.44$\pm$0.36&0.303$\pm$0.022&8.604$\pm$0.014\\
\hline\hline
\multirow{9}{*}{2.09--2.61} & \multirow{3}{*}{Field}
&53&36&9.77$\pm$0.02&0.92$\pm$0.16&0.090$\pm$0.009&8.305$\pm$0.023\\ &
&36&27&10.22$\pm$0.02&1.27$\pm$0.16&0.180$\pm$0.011&8.476$\pm$0.015\\ &
&7&7&10.65$\pm$0.06&1.09$\pm$0.24&0.308$\pm$0.031&8.609$\pm$0.025\\

\cline{2-8} 
                  & \multirow{3}{*}{Intermediate-density} 
&42&36&9.76$\pm$0.02&0.91$\pm$0.14&0.084$\pm$0.008&8.288$\pm$0.022\\ &
&42&27&10.21$\pm$0.02&1.28$\pm$0.16&0.165$\pm$0.013&8.454$\pm$0.020\\ &
&9&7&10.66$\pm$0.02&1.38$\pm$0.16&0.219$\pm$0.020&8.524$\pm$0.023\\

                 \cline{2-8} 
                  & \multirow{3}{*}{Overdense} &45&36&9.77$\pm$0.02&0.99$\pm$0.13&0.072$\pm$0.011&8.249$\pm$0.036\\ &
&39&27&10.21$\pm$0.02&1.31$\pm$0.14&0.144$\pm$0.014&8.420$\pm$0.024\\ &
&21&7&10.66$\pm$0.03&1.51$\pm$0.39&0.209$\pm$0.038&8.512$\pm$0.024\\

                 \hline
\end{tabular}%
\end{table*}

Analysis of the mass-controlled samples confirms the trends already observed for the unmatched sample in Figure \ref{fig:MZR_fit}; however, the significance of observed trends is more reliable when we control the stellar mass distribution of the sample. Considering all stellar mass ranges ($\rm 10^{9.5} M_\odot \lesssim M_* \lesssim 10^{11} M_\odot$), at $z\sim 1.5$, on average galaxies in overdensities are $\sim 0.07$ dex rich in metal compared to their field counterparts. In contrast, at $z\sim 2.3$, average galaxies in overdensities have $\sim 0.07$ dex lower metallicities compared to the field galaxies. Before discussing the physical origin of the observed trends in detail, we compare our results with previous works in the following section. 

\section{Discussion}
\label{sec:Discussion}
\subsection{Comparison with previous works}
\label{Comparison with previous works}
Only a handful of studies have been conducted to investigate the role of the environment in the MZR at $z>1$. The bottom panels in Figure \ref{MZR_mass_controlled} show the offset between the average metallicity of galaxies in overdensities (protocluster/cluster) and that of field galaxies as a function of stellar mass at both redshift bins, $z\sim 1.5$ and 2.3. In these figures, we also include the aforementioned offsets from literature. Our finding at $z\sim 1.5$ is in full agreement with the recent work of \cite{Maier19}. They studied a massive cluster at $z\sim 1.5$ and found that, at a given stellar mass, the metallicities of galaxies in the inner part of the cluster are higher by $\sim 0.1$ dex than those of infalling and field galaxies. They suggest that strangulation in the dense cores of clusters results in a cold gas removal that enhances the metallicity. However, \cite{Namiki19} and \cite{Tran15} found no significant environmental dependence of the MZR around $z\sim 1.5$. \cite{Namiki19} compared the MZR of narrow-band H$\alpha$-selected cluster members with the field MZR of \cite{Stott13}. \cite{Tran15} considered the field sample of \cite{Zahid14} for the comparison. Thus, we also utilize the field samples of \cite{Stott13} and \cite{Zahid14} to measure the metallicity offset (shown in Figure \ref{MZR_mass_controlled}) for the studies of \cite{Namiki19} and \cite{Tran15}, respectively.       

In the studies regarding the environmental dependence of the MZR, the sample selection bias needs to be handled properly to ensure that both field and cluster samples are selected in the same way. For example, \cite{Stott13} found that H$\alpha$ emitting galaxies selected from High-Z Emission Line Survey (HiZELS) \citep{Sobral13} have remarkably higher metallicity at a given stellar mass compared to a rest-frame UV-selected sample of \cite{Erb2006}. They argued that the UV-selected sample of \cite{Erb2006} tends to have a higher average SFR compared to the HiZELS narrow-band selected sample, resulting in a bias against metal-rich galaxies. Therefore, the comparison between the MZR of two different works can be biased due to the selection criteria. However, this is not an issue when the sample is selected uniformly in different environments, and measurements are performed consistently. \cite{Shimakawa15} studied narrow-band selected galaxies in two rich overdensities at $z=2.2$ and 2.5. They found that the metallicity of protocluster galaxies ($\rm M_*<10^{11}\ M_\odot$) is $\sim 0.15$ dex higher than that of field galaxies, which contrasts with our result at $z\sim 2.3$. According to \cite{Stott13} and discussions included in \cite{Shimakawa15}, selection bias is a potential concern in their study since different criteria are used for the selection of protocluster (narrow-band selected) and field \cite[UV-selected,][]{Erb2006} galaxies. This concern can be addressed in future studies by comparing the protocluster MZR of \cite{Shimakawa15} with a field MZR of a narrow-band selected H$\alpha$ emitters at $z\sim 2.2$.

\cite{Kacprzak15} and \cite{Alcorn09} studied the MZR of a protocluster \citep{Yuan14} at $z=2.1$ and reported no significant environmental effect on the MZR. As shown in Figure \ref{Density map}, we also captured this protocluster in the present work. We speculate that the lack of environmental dependence of the MZR in their studies can originate from considering all the protocluster members at the same density contrast. As seen in Figure \ref{Density map}, the protocluster includes different components and considering that all the members are located in an overdensity weakens any existing environmental dependence of the MZR.

\cite{Valentino15} found that cluster star-forming galaxies ($\rm 10^{10}M_\odot\leq M_*\leq10^{11}M_\odot $) are $\sim 0.25$ dex poorer in metals than their field counterparts at $z\sim 2$. We find a similar trend at $z\sim 2.3$ with a lower average metal deficiency of $\sim 0.1$ dex. Moreover, Sattari et al. {\it in prep.} studied a massive protocluster at $z=2.2$ \citep{Darvish20} and found that protocluster galaxies with $\rm M_*\sim10^{9.6}\ M_\odot$ and $\rm M_*\sim10^{10.2}\ M_\odot$ are $0.03\pm 0.06$ and $0.10 \pm 0.04$ dex metal deficient, respectively, compared to field galaxies. Our findings at $z\sim 2.3$ are in agreement with their results.      

In the presence of limited sample size, unbalanced stellar mass distribution of galaxies between cluster members and the field sample can affect the environmental trends. In other words, even in each stellar mass bin (e.g., $\rm 9.5\leq\log(M_*/M\odot)< 10$), the stacked spectra are biased toward massive galaxies, which are usually detected with higher S/N. Therefore, one needs to match the shape of stellar mass distributions in different environments as described in Section \ref{Mass-controlled sample}, and having the same average/median stellar mass in the bins of the environment does not guarantee that the effect of stellar mass on gas-phase metallicity is removed properly.  \cite{Kulas13} have studied the MZR of a protocluster at $z\sim 2.3$ and reported a $0.1$ dex metallicity enhancement with respect to the field galaxies. We do not confirm such trends in the present work. Their field and protocluster samples are selected consistently (UV-selected). However, based on Figure 2 of \cite{Kulas13}, we speculate that the metallicity enhancement seen in their work can be affected by the significantly different stellar mass distributions of their protocluster and field samples. 

Beyond the local Universe, all the previous works regarding environmental dependence of gas-phase metallicity are conducted by comparing MZR for a cluster/protocluster and a sample of field galaxies. However, within a given cluster/protocluster, galaxies may have different density contrasts. Thus, quantifying the environment using local density is a better approach to study environmental effect rather than considering all the members of a cluster/protocluster residing in an overdensity. For instance, \cite{Maier19} found that the metallicity enhancement at $z\sim 1.5$ is only observed for the core (inside half of $\rm R_{200}$) of a cluster, not for infalling protocluster members. Assuming that all the cluster/protocluster members have similar local densities weakens any underlying environmental dependence of the MZR. In the present work, for the first time, we study the gas-phase metallicity of galaxies as a function of their local density and its evolution with cosmic time beyond the local Universe. Similar works have been conducted by \cite{Mouhcine07,Cooper08,Peng14} with SDSS sample at $z\sim 0$.

\subsection{How does the environment affect MZR?}
\label{How does environment affect MZR?}

In this work, we find that galaxies in overdense regions have lower metallicity than their field counterparts at $z\sim 2.3$, but they become more metal-rich as they evolve to $z\sim 1.5$. In other words, the gas-phase metallicity of galaxies in a dense environment increases by $\sim 0.15$ dex as they evolve from $z\sim 2.3$ to $z\sim 1.5$ ($\sim 1.5$ Gyr), but the metallicity of field galaxies are almost unchanged over this period. It implies that, at high redshift, metal enrichment processes are affected by the environment where galaxies reside.

Previous studies observed that dense environments at the early stage of galaxy cluster formation ($z\gtrsim 2$) contain a significant fraction of pristine gas \citep[e.g.,][]{ Cucciati14}. In the absence of gravitational heating processes, gas accretion in overdensities should be more prominent due to their deep potential well. However, infalling gas in overdensities with massive halos gets shock-heated and needs to radiate its kinetic energy to accrete into the halo. At high redshifts, where the average density of the Universe is higher by a factor of $(1+z)^3$, gas cooling is very efficient \citep{vandeVoort2012}. It is also observed that overdense regions in the early Universe are not only overdense in galaxies but also contain a large fraction of dense gas \citep[e.g.,][]{Hennawi15}. Denser gas can cool down faster, facilitating the accretion of the metal-poor primordial gas into galaxies \citep{Kere2005,Dekel2006}. As a result, the prominent accretion of cold metal-poor gas in overdensities dilutes the metal content of ISM in galaxies at high redshifts. This results in the metal-poor gas in the galaxies residing in dense environments at high redshifts, as observed in the present work at $z\sim 2.3$. \cite{Valentino15} also observed the same metal deficiency at $z\sim 2$ and concluded that the accretion of pristine gas from cluster-scale reservoirs lowers the gas-phase metallicity of galaxies in dense environments compared to their coeval field galaxies.

In contrast, at the lower redshift, $z\lesssim 2$, the gas cannot cool down efficiently in overdensities, so the galaxies in those regions start to experience cosmological starvation. The lack of pristine metal-poor gas accretion in cluster members has been observed in different simulations out to $z\sim 2$ \citep{vandeVooet2017,Gupta18}. Moreover, \cite{Gupta18} studied chemical pre-processing of cluster galaxies in the IllustrisTNG cosmological simulation and found that at $z=1.5$, cluster galaxies receive $\sim 0.05$ dex more metal-rich infalling gas than galaxies in the field. But, this metallicity enhancement disappears at higher redshifts ($z> 1.5$). At $z\sim 2$, when galaxies actively form their stars, feedback processes should be strong enough to expel part of the processed gas into the IGM through the outflows. As a result, crowded regions contain pre-processed and metal-enriched gas, which can be then re-accreted to the galaxies at lower redshifts. As pre-processed gas has higher metallicity, it can cool down faster which facilitates its accretion \citep{Kere2005,Dekel2006}. Therefore, we speculate that both effects, suppressed primordial metal-poor gas infall and pre-processed metal-enriched gas accretion, are essential in ramping up metal production in a dense environment around $z\lesssim 2$. This can explain our result at $z\sim 1.5$, where we find metallicity enhancement for galaxies in overdensities compared to field galaxies.

In the absence of cold gas accretion, galaxies could maintain their SFR unchanged for a period of time as they start to consume their gas reservoirs. This time ranges from a few hundred Myr for most massive galaxies up to a few Gyr for low-mass galaxies \citep{McGee2014,Balogh2016}. However, 
the dilution of ISM's metal content will be ceased immediately after the termination of cold gas accretion. Therefore, the absence of SFR suppression in overdense regions at $z\sim 1.5$ for our star-forming sample does not contradict the observed metal enhancement in galaxies residing in overdensities at that redshift. Given the relatively short cosmic time interval between $z\sim 2.3$ and $z\sim 1.5$ ($\sim 1.5\ \rm Gyr$), galaxies with stellar mass range probed in the present work ($\rm M_*<10^{11} M_\odot$) do not have enough time to consume their remaining gas reservoirs after the halt of cold gas accretion. 

\cite{Sanders18} showed that the MZR varies with SFR at $z\sim 2.3$, such that the gas-phase metallicity of galaxies at fixed $\rm M_*$ is anticorrelated with their SFRs.   
Using SED-derived SFRs, we also find slight evidence of enhanced SFR for star-forming galaxies located in overdensities. Therefore, the prominent gas accretion in overdense regions at high redshifts can explain both lower gas-phase metallicity and higher SFR of galaxies residing in overdensities at $z\sim 2.3$. It is worth noting that gas outflows can also play a significant role in lowering the gas-phase metallicity of galaxies located in overdensities and actively forming stars. These outflows can re-accrete into galaxies at lower redshifts and increase the gas-phase metallicity of galaxies as found in the present work at $z\sim 1.5$.


\section{Summary}
\label{sec:Summary}
Using a large near-IR spectroscopic sample drawn from the MOSDEF survey, combined with the local density measurements from the CANDELS photometric survey, we study the environmental dependence of the MZR at $z\sim 1.5$ and $z\sim 2.3$. We cross-match MOSDEF galaxies with the publicly available catalog of local density measurements in five CANDELS fields \citep{chartab19}, and use the $\rm N2=\frac{[NII]\lambda6584}{H\alpha}$ indicator to measure the gas-phase oxygen abundances of 167 galaxies at $1.37\leq z\leq1.7$ and 303 galaxies at $2.09\leq z\leq2.61$.

The samples are labeled as overdense, intermediate-density, and field based on their local density measurements. We match the stellar mass distribution of our sample in three different environments to properly disentangle the effects of stellar mass from those related to the environment. Massive galaxies are mostly found in overdensities and unmatched underlying stellar mass distributions between different environments can affect the strength of the trends or even change the observed trends in the presence of a limited sample size. For the mass-matched sample, our findings can be summarized as follows:

\begin{itemize}
    \item At $z\sim 1.5$, the average metallicity of galaxies in overdensities with $ \rm M_*\sim 10^{9.8} M_\odot, 10^{10.2} M_\odot$ and $\rm 10^{10.8} M_\odot$ is higher relative to their field counterparts by $0.094\pm 0.051$ (1.8$\sigma$ significance), $0.068\pm 0.028$ (2.4$\sigma$ significance) and $0.052\pm 0.043$ (1.2$\sigma$ significance) dex, respectively. Also, the metallicity enhancements for $\rm M_*\sim 10^{9.8} M_\odot$ and $\rm 10^{10.8} M_\odot$ galaxies in intermediate-densities are $0.090\pm 0.052$ (1.7$\sigma$ significance) and $0.077\pm 0.042$ (1.8$\sigma$ significance) dex, respectively, with being insignificant for $\rm M_*\sim \rm 10^{10.2} M_\odot$ galaxies.
 
    \item At $z\sim 2.3$, galaxies that reside in overdensities with $ \rm M_*\sim 10^{9.8} M_\odot, 10^{10.2} M_\odot$ and $\rm 10^{10.7} M_\odot$, have lower gas-phase metallicity by $0.056\pm 0.043$ (1.3$\sigma$ significance), $0.056\pm 0.028$ (2$\sigma$ significance) and $0.096\pm 0.034$ (2.8$\sigma$ significance) dex compared to their coeval field sample, respectively. This metal deficiency is insignificant for galaxies residing in intermediate-densities except for the massive galaxies ($\rm M_*\sim \rm 10^{10.7} M_\odot$), where we found $0.085\pm 0.034$ (2.5$\sigma$ significance) dex metal deficiency compared to field counterparts.
    
    \item Our results suggest that the efficient gas cooling mechanisms at high redshifts result in the prominent accretion of primordial metal-poor gas into the galaxies in overdensities. This cold metal-poor gas can dilute the metal content of ISM gas and lowers the gas-phase metallicity of galaxies, as seen in the present work ($z\sim 2.3$). However, as galaxies evolve to the lower redshifts ($z\lesssim 2$), the shock-heated gas in overdensities with massive halos cannot cool down efficiently, which prevents it from accreting into the galaxy. The termination of pristine gas accretion in overdensities along with the accretion of pre-processed gas due to the strong outflows increase the metallicity of galaxies at lower redshifts, $z<2$. This scenario can explain our result at $z\sim 1.5$, where we find metallicity enhancement for galaxies in overdensities compared to coeval field galaxies.
\end{itemize}

\section*{Acknowledgments} 
We thank the anonymous referee for providing insightful comments and suggestions that improved the quality of this work. We acknowledge support from NSF AAG grants AST-1312780, 1312547, 1312764, and 1313171, archival grant AR-13907 provided by NASA through the Space Telescope Science Institute. I.S. is supported by NASA through the NASA Hubble Fellowship grant \# HST-HF2-51420, awarded by the Space Telescope Science Institute, which is operated by the Association of Universities for Research in Astronomy, Inc., for NASA, under contract NAS5-26555. The authors wish to extend special thanks to those of Hawaiian ancestry on whose sacred mountain we are privileged to be guests. Without their generous hospitality, most of the observations presented herein would not have been possible.

\bibliography{environment_MOSDEF}

\appendix
\counterwithin{figure}{section}
\section{The MZR with $\rm [NII]\lambda 6584$-detection requirement}
\label{appendix}
\begin{figure*}[h]
    \centering
    \subfloat{{\includegraphics[width=8.5cm, trim=0.6cm 0.1cm 1.5cm 0.75cm]{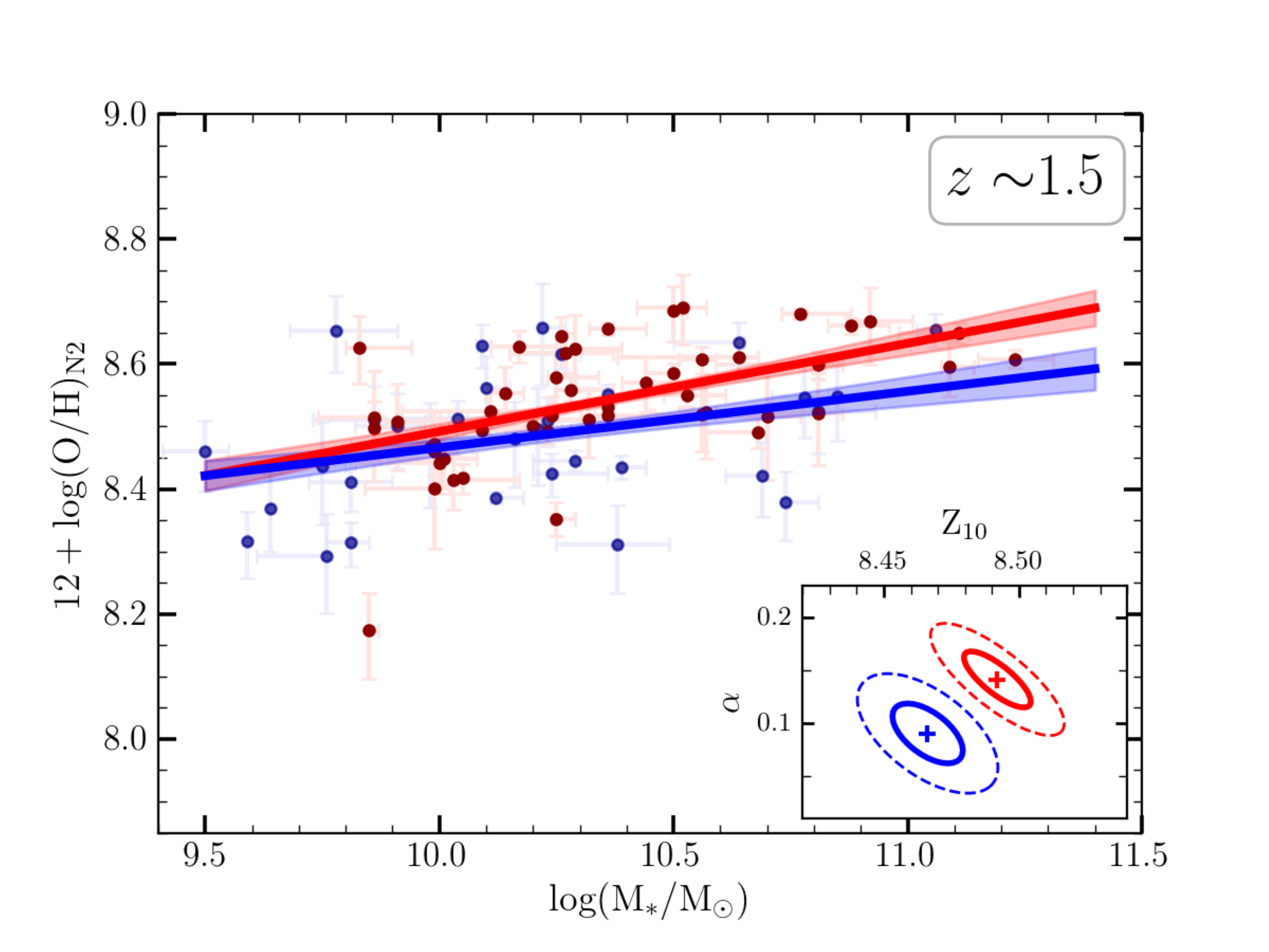} }}%
    \qquad
    \subfloat{{\includegraphics[width=8.5cm, trim=0.6cm 0.1cm 1.5cm 0.75cm]{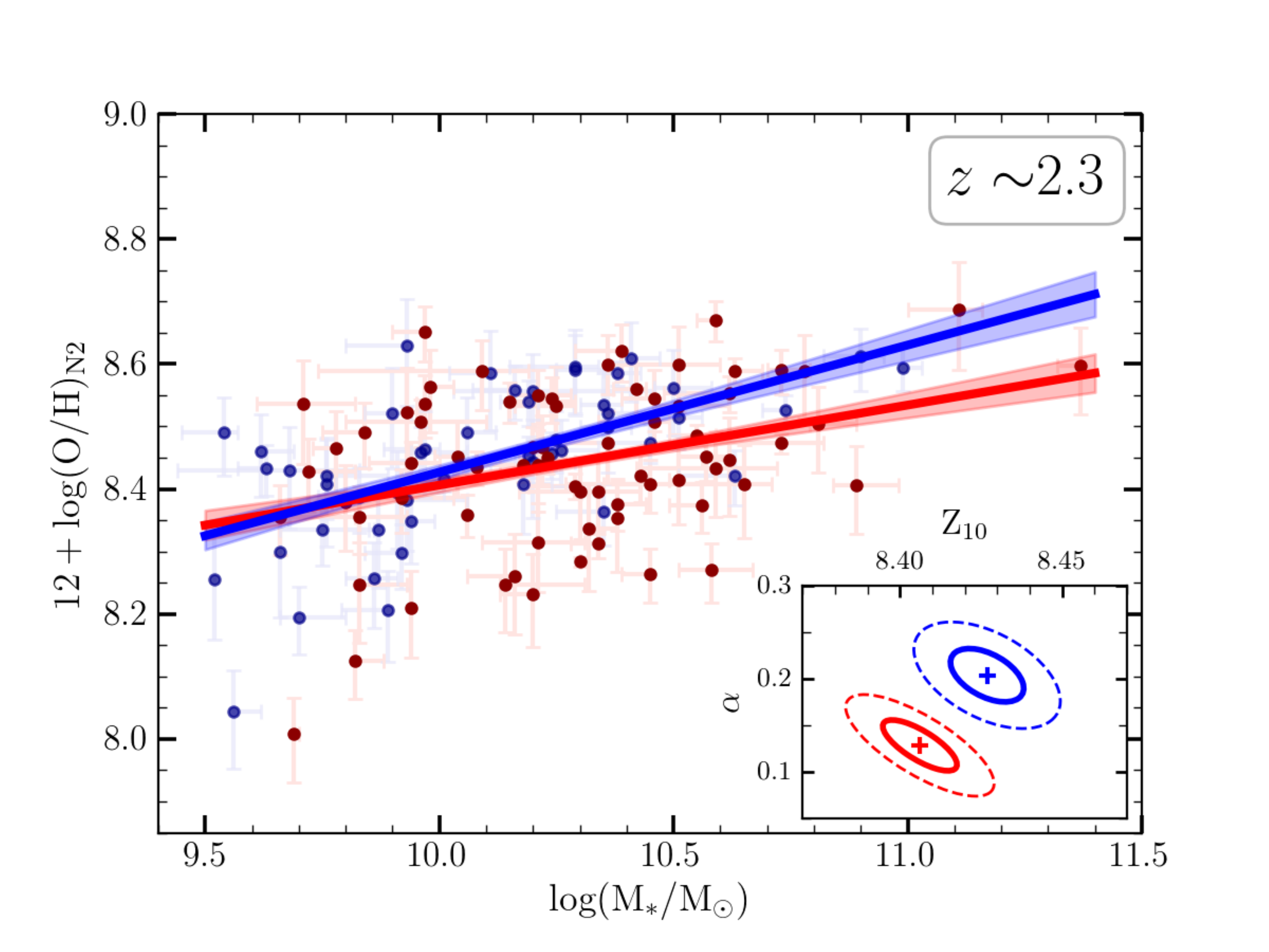} }}%

    \caption{Similar to Figure \ref{fig:MZR_fit}, but here we ignore $\rm [NII]\lambda 6584$ non-detection galaxies. The best-fit lines to MZRs are shown for galaxies with $\rm S/N>3$ detection in $\rm [NII]\lambda 6584$, residing in two extreme environments, overdensity (red) and underdensity (blue). The shaded regions around the best-fit models show $1\sigma$ error in the best-fit lines. Both gas-phase metallicity and stellar-mass errors are taken into account in regression analysis. Sub-panels show the best value (“+”), $1\sigma$ (solid) and $2\sigma$ (dashed) confidence intervals for the two-dimensional posterior distribution of the slope and the intercept, which are considered to be free parameters of the linear model, $\rm12+\log(O/H)= Z_0+\alpha [\log({M_*}/{M_\odot})-10]$.}%
    \label{fig:MZR_fit_non_stack}%
\end{figure*}

We perform linear regression for individual galaxies (non-stacked) to calculate the best-fit MZR for galaxies with significant detection ($\rm S/N>3$) in $\rm [NII]\lambda 6584$. We note that requiring detection in $\rm [NII]\lambda 6584$ introduces a bias to our sample toward higher gas-phase metallicities, especially in the low-mass end of the MZR where non-detections are prevalent; however, it is worth investigating the offset between the best-fit MZRs in different environments without considering the contribution of $\rm [NII]\lambda 6584$ non-detections. We fit a linear model, $\rm12+\log(O/H)= Z_0+\alpha [\log({M_*}/{M_\odot})-10]$, to the MZR of the samples in two extreme environment bins (underdensity and overdensity) considering both measurement errors in stellar mass and gas-phase metallicity. Figure \ref{fig:MZR_fit_non_stack} shows the best-fit lines along with 2D-posterior distributions of the slope and the intercept for the bins of environment. A significant distinction ($>2\sigma$) between the posterior of fit parameters for the field galaxies and those located in overdensities suggests that the metallicity is enhanced at $z\sim 1.5$ and suppressed at $z\sim 2.3$ for galaxies in overdensities compared to field counterparts. Based on the best-fit models and their corresponding uncertainties, on average, galaxies in overdensities at $z\sim 1.5$ have $0.050\pm0.024$ dex higher gas-phase metallicity compared to coeval field galaxies. The trend reverses at higher redshift, $z\sim 2.3$, such that galaxies residing in overdensities are metal deficient by $0.055\pm0.025$ dex than the field counterparts. These results are in general agreement with our findings using mass-matched stacked spectra (Figure \ref{MZR_mass_controlled}). Moreover, we estimate the intrinsic scatter of the MZR ($\sigma_{\rm int}$) in both extreme environments at $z\sim 1.5$ and $z\sim 2.3$. We assume that the observed scatter ($\sigma_{\rm obs}$) around the best-fit MZR is $\sigma_{\rm obs}^2=\sigma_{\rm int}^2+\sigma_{\rm meas}^2$, where $\sigma_{\rm meas}$ is the average measurement uncertainty. We estimate that the intrinsic scatter of the MZR at $z\sim 1.5$ ($z\sim 2.3$) is $0.07\ (0.07)$ and $0.07\ (0.10)$ dex for the field galaxies and those residing in overdensities, respectively, which are consistent with the intrinsic scatter of $z\sim 0$ MZR in different environments \citep{Cooper08}. We note that, although ignoring $\rm [NII]\lambda 6584$ non-detections does not change our conclusions, the bias is evident in the low-mass end of the MZR by comparing best-fit lines in Figure \ref{fig:MZR_fit_non_stack} with gas-phase metallicities derived from mass-matched stacked spectra (Figure \ref{MZR_mass_controlled}). Therefore, the stacking technique employed in Section \ref{environment_MZR} is the preferred method as it provides an unbiased MZR where the contribution of $\rm [NII]\lambda 6584$ non-detections are taken into account properly.

\end{document}